\documentclass[preprint]{elsarticle}

\usepackage{hyperref}
\usepackage{amssymb}

\begin{document}

\title{On the concordance of cosmological data in the case of 
the generalized Chaplygin gas}

\author{R.\ Aurich and S.\ Lustig}


\address{Institut f\"ur Theoretische Physik, Universit\"at Ulm,\\
Albert-Einstein-Allee 11, D-89069 Ulm, Germany}

\begin{abstract}
The generalized Chaplygin gas cosmology provides a prime example for the
class of unified dark matter models,
which substitute the two dark components of the standard cosmological
$\Lambda$CDM concordance model by a single dark component.
The equation of state of the generalized Chaplygin gas is characterised by a
parameter $\alpha$ such that the standard $\Lambda$CDM model is recovered
in the case $\alpha=0$ with respect to the background dynamics and the
cosmic microwave background (CMB) statistics.
This allows to investigate the concordance of different cosmological data sets
with respect to $\alpha$.
We compare the supernova data of the Supernova Cosmology Project,
the data of the baryon oscillation spectroscopic survey (BOSS) of
the third Sloan digital sky survey (SDSS-III) and the
CMB data of the Planck 2015 data release.
The importance of the BOSS Lyman $\alpha$ forest BAO measurements
is investigated.
It is found that these data sets possess a common overlap of the
confidence domains only for Chaplygin gas cosmologies very close to the
$\Lambda$CDM model.
\end{abstract}

\begin{keyword}
dark energy theory \sep cosmic microwave background \sep large-scale structure
\PACS 98.80.-k \sep 98.70.Vc \sep 98.80.Es
\end{keyword}

\maketitle

\biboptions{compress}

\section{Introduction}
\label{sec:intro}

For almost two decades, cosmology possesses a standard cosmological model
which allows a remarkable successful description of a large variety
of observational data, the $\Lambda$CDM concordance model.
For a recent review on the various data sets, see e.\,g.\
\cite{Huterer_Shafer_2017}.
Although there is currently no competing cosmological model,
there remain tensions \cite{Buchert_Coley_Kleinert_Roukema_Wiltshire_2016}
which justify the investigation of alternatives.
The $\Lambda$CDM concordance model is based on two dark ingredients,
the dark energy in the form of the cosmological constant $\Lambda$ and
the cold dark matter (CDM).
There have ever been attempts to establish alternative cosmological models
which are based only on a single dark component,
the so-called unified dark matter (UDM) models.

The prototypical model for a UDM cosmology is provided by a dark matter fluid
having the equation of state $p = -A/\varepsilon$
of the Chaplygin gas,
where $\varepsilon$ and $p$ denote the energy density and the pressure,
respectively, and $A$ is a model parameter.
The Chaplygin gas was put in a cosmological context by the work of
Kamenshchik, Moschella and Pasquier \cite{Kamenshchik_Moschella_Pasquier_2001}
where also an obvious generalisation
of the equation of state $p = -A/\varepsilon^\alpha$ with $\alpha > -1$
was introduced.
For some further early works on the Chaplygin gas cosmology,
see \cite{Fabris_Goncalves_deSouza__2001,%
Bilic_Tupper_Viollier_2002,%
Bento_Bertolami_Sen_2002,Bento_Bertolami_Sen_2003,%
Bean_Dore_2003,%
Amendola_Finelli_Burigana_Carturan_2003,%
Carturan_Finelli_2003,%
Makler_deOliveira_Waga_2003}.
The paper \cite{Dev_Alcaniz_Jain_2003} tests the Chaplygin gas against
gravitational lenses and age estimations.
The generalised Chaplygin model has the nice property that the
background cosmology is for $\alpha=0$ identical to the $\Lambda$CDM model.
This allows to test how far the equation of state of the dark fluid
can deviate from that of the cosmological constant $\Lambda$,
that is $p = -\varepsilon$.
Then, it was discovered \cite{Sandvik_Tegmark_Zaldarriaga_2004}
that a wide class of UDM cosmologies suffer from instabilities
which lead to matter power spectra incompatible with the observations.
Especially, it was found in \cite{Sandvik_Tegmark_Zaldarriaga_2004}
that the generalised Chaplygin gas cosmologies with $|\alpha| \gg 10^{-5}$
are ruled out.
Thus, only a minute parameter space of the generalised Chaplygin cosmology
was left with values of $\alpha$ so close to zero
that these models are almost indistinguishable from the $\Lambda$CDM model.

A loophole for the UDM models is already proposed in
\cite{Sandvik_Tegmark_Zaldarriaga_2004}.
It is noted that the dark fluid considered so far does not
exploit the full range of physical fluid properties.
It is shown in \cite{Hu_1998} that the fluid is not fully specified by
defining the adiabatic speed of sound $c_{\hbox{\scriptsize ad}}$,
but in addition, varying the effective speed of sound $c_{\hbox{\scriptsize eff}}$
and the parameter $c_{\hbox{\scriptsize vis}}$, as defined in \cite{Hu_1998},
leads to a plethora of behaviours.
It is shown in \cite{Reis_Waga_Calvao_Joras_2003}
that choosing the effective speed of sound $c_{\hbox{\scriptsize eff}}$
different from the adiabatic one leads to an admissible parameter range
of $\alpha$ that is significantly extended.
Thus a large class of generalised Chaplygin cosmologies is compatible with
the large scale structure and CMB observations,
if $c_{\hbox{\scriptsize eff}} \neq c_{\hbox{\scriptsize ad}}$ is allowed.

This generated again a large interest in the Chaplygin cosmologies
and its variants, see e.\,g.\
\cite{Nozari_Azizi_Alipour_2011,%
Lamon_Woehr_2010,%
Roy_Buchert_2010,%
Bento_Bertolami_Reboucas_Silva_2006,%
Wu_Yu_2007,%
Xu_Lu_2010,%
Campos_Fabris_Perez_Piattella_2013,%
Xu_Zhang_2016,%
Freitas_Goncalves_Velten_2011,%
El_Zant_2015,%
Lu_Xu_Wu_Liu_2011,%
Paul_Thakur_Beesham_2014,%
Zhang_Wu_Zhang_2006,%
Pourhassan_Kahya_2014,%
Carneiro_Pigozzo_2014,%
Marttens_etc_2017,%
Wen_Wang_2017}.
Most papers assume a flat universe when the allowed parameter space is
estimated according to various cosmological data.
Assuming a flat universe, the $2\sigma$ estimation $\alpha=-0.14_{-0.19}^{+0.30}$
is stated in \cite{Wu_Yu_2007} from a joint analysis of
supernovae, BAO, and CMB data.
A recent analysis, which also assumes a flat universe,
finds the $1\sigma$ estimation $\alpha=-0.03_{-0.057}^{+0.067}$
\cite{Xu_Zhang_2016}.
Both use for their CMB analysis a compressed likelihood
which reduces the CMB information to a few numbers
among them the shift parameter ${\cal R}$
\cite{Bond_Efstathiou_Tegmark_1997,Efstathiou_Bond_1999}.
However, since the matter-like content is not conserved in UDM cosmologies,
the shift parameter ${\cal R}$ is not strictly applicable in this case.
Negative values for $\alpha$ are also found for a flat universe by including
gamma-ray bursts data \cite{Freitas_Goncalves_Velten_2011},
however, the reliability of these data is strongly debated.
To be compatible with the rotation curves of galaxies,
dark matter haloes consisting of pure generalised Chaplygin gas
should satisfy $\alpha\lesssim 0.0001$ \cite{El_Zant_2015}.

As a modification of the generalised Chaplygin gas,
the equation of state $p=\beta\rho-A/\rho^\alpha$ is proposed,
but it is found in \cite{Lu_Xu_Wu_Liu_2011,Paul_Thakur_Beesham_2014}
that $\beta$ has to be very small.
Thus, both the energy density and the equation of state approach those
of the generalised Chaplygin gas.
Also not exactly comparable, their values for $\alpha$,
that is $\alpha =0.11_{-0.25}^{+0.34}$ \cite{Lu_Xu_Wu_Liu_2011} and
the $1\sigma$ interval $\alpha\in [0.08,0.31]$ \cite{Paul_Thakur_Beesham_2014},
should be of the order of those expected in the generalised Chaplygin gas.
Their CMB analysis also makes use of the shift parameter ${\cal R}$.

The generalised Chaplygin gas can also serve as a model for
interacting dark matter \cite{Zhang_Wu_Zhang_2006}.
Using supernovae data calibrated with different fitters in a joint analysis,
even smaller values of $\alpha$ are found in \cite{Carneiro_Pigozzo_2014},
such as $\alpha\simeq -0.5$ or $\alpha\simeq -0.36$,
which shows the sensitivity of the chosen calibration method of SNe Ia data.
Using additional data sets, a joint analysis carried out in
\cite{Xu_Zhang_2016} finds the $1\sigma$ estimation
$\alpha=0.004_{-0.010}^{+0.013}$.
This analysis assumes a flat universe, which is also the case for the recent
joint analysis \cite{Marttens_etc_2017} which states $|\alpha|<0.05$.
Although \cite{Marttens_etc_2017} uses the full CMB spectra, their analysis
does not vary all cosmological parameters,
so that it is not a best-fit analysis.
Furthermore, they use the binned JLA supernova data
\cite{Betoule_et_al_2014},
which are strongly dependent on the fiducial $\Lambda$CDM model.

It is worthwhile to remark that the correspondence between the
generalised Chaplygin gas cosmology for $\alpha=0$ and the
$\Lambda$CDM cosmology refers to the background level.
On the perturbation level, there are differences but it turns out
that the CMB statistics is the same within linear perturbation theory.
Since we consider only background data and CMB data in this paper,
we equate the generalised Chaplygin gas cosmology for $\alpha=0$
with the $\Lambda$CDM model,
but the difference should be kept in mind.

So there is a wide range of possible parameter values and structures
for the equation of state.
It is the aim of this paper to provide a parameter estimation of the
generalised Chaplygin gas cosmology without the restriction to
flat universes.
Furthermore, the tensions between the different data sets
with respect to the Chaplygin gas cosmology are discussed.
A special focus is put on the question for which equation of state of
the generalised Chaplygin gas a concordance between the supernovae,
BAO, and CMB data is achieved.
In section \ref{sec:Chaplygin} the Chaplygin cosmology is introduced,
while section \ref{sec:data} states the details about the cosmological data
used for the parameter estimation.
Section \ref{sec:Analysis_and_Results} presents the analysis and
section \ref{sec:Summary} concludes with a summary of the results.

\section{The generalised Chaplygin gas cosmology}
\label{sec:Chaplygin}

The equation of state of the generalised Chaplygin gas is defined as
\begin{equation}
\label{Eq:Def_eos_Chaplygin}
p \; = \; -\, \frac{A}{\varepsilon^\alpha}
\end{equation}
with $\alpha>-1$ and $A$ a constant parameter.
The original Chaplygin gas is obtained for $\alpha=1$.
The integration of the continuity equation leads to the energy density
$\varepsilon_{\hbox{\scriptsize Chap}}(z)$ of the generalised Chaplygin gas
as a function of the redshift $z$
\begin{equation}
\label{Eq:Chaplygin_energy_density}
\varepsilon_{\hbox{\scriptsize Chap}}(z) \; = \;
\varepsilon_{\hbox{\scriptsize Chap}}^0 \, \Big\{
\left(1 - B\right) \, (z+1)^{3(\alpha+1)} + B \Big\}^{1/(\alpha+1)}
\hspace{4pt} \hbox{ with } \hspace{4pt}
B := \frac A{(\varepsilon_{\hbox{\scriptsize Chap}}^0)^{\alpha+1}}
\hspace{4pt} .
\end{equation}
This leads to the equation of state
\begin{equation}
\label{Eq:Chaplygin_equation_of_state}
w_{\hbox{\scriptsize Chap}}(z) \; = \; \frac p\varepsilon \; = \;
-  \, \frac{B}{
\left(1 - B\right) \, (z+1)^{3(\alpha+1)} + B}
\hspace{10pt} ,
\end{equation}
which satisfies one of the most important properties of a UDM model:
It has to behave matter-like at an early epoch in the history
of the universe in order to allow sufficient structure formation,
and dark-energy-like at later times in order to explain
the accelerated expansion.
The generalised Chaplygin gas model has the remarkable property
that its  background model is for $\alpha=0$ and 
$B=1/(1+\Omega_{\hbox{\scriptsize CDM}}/\Omega_\Lambda)$
identical to that of the $\Lambda$CDM concordance model with the same
parameters $\Omega_{\hbox{\scriptsize CDM}}$ and $\Omega_\Lambda$.

In our analysis we will exclude the parameter space with $B>1$,
since the present equation of state is then $w_{\hbox{\scriptsize Chap}}(z=0)=-B<-1$
which corresponds to the so-called phantom energy.
But more worse, at the redshift $z=(B/(B-1))^{1/3(\alpha+1)}-1>0$
the equation of state becomes singular.
It will turn out that this restriction is only relevant,
if one uses solely the BAO data with $z<1$ for matching the
Chaplygin cosmology.

The effective content of cold dark matter of a UDM model can be computed
as argued in \cite{Aurich_Lustig_2016} by rewriting the evolution of the
dark energy density as
\begin{eqnarray} \nonumber
\varepsilon_{\hbox{\scriptsize de}}(z) & = &
\varepsilon_{\hbox{\scriptsize de}}^0 \, \exp\left( 3 \int_{x(z)}^1
\frac{1+w_{\hbox{\scriptsize de}}(x)}x\, dx \right) \\
& = &
\label{Eq:Def_epsilon_de}
\varepsilon_{\hbox{\scriptsize de}}^{\hbox{\scriptsize eff}} \,
\left(\frac{a_0}{a(z)}\right)^3
\, \exp\left(-3 \int_0^{x(z)} \frac{w_{\hbox{\scriptsize de}}(x)}x\, dx \right)
\hspace{10pt} ,
\end{eqnarray}
where $x:= a(z)/a_0$ is the scale factor $a(z)$ normalised to one
at the present time and
$\varepsilon_{\hbox{\scriptsize de}}^0$ is the current energy density.
Here, we have defined
\begin{equation}
\label{Def:epsilon_de_eff}
\varepsilon_{\hbox{\scriptsize de}}^{\hbox{\scriptsize eff}} \; := \;
\varepsilon_{\hbox{\scriptsize de}}^0 \, \exp\left( 3 \int_0^1
\frac{w_{\hbox{\scriptsize de}}(x)}x\, dx \right)
\hspace{10pt} ,
\end{equation}
which converges for an equation of state with $w_{\hbox{\scriptsize de}}(x)\to 0$
for $x\to 0$ as it is the case for UDM models.
The quantity $\varepsilon_{\hbox{\scriptsize de}}^{\hbox{\scriptsize eff}}$ measures
the effective content of cold matter at early times,
since the last exponential factor in eq.\,(\ref{Eq:Def_epsilon_de})
takes on values close to one at early times.

It is convenient to define the cosmological parameters
\begin{equation}
\label{Def:Omega_de_eff}
\Omega_{\hbox{\scriptsize de}} \; := \;
\frac{8\pi G}{3 H_0^2c^2} \, \varepsilon_{\hbox{\scriptsize de}}^0
\hspace{10pt} \hbox{ and } \hspace{10pt}
\Omega_{\hbox{\scriptsize de}}^{\hbox{\scriptsize eff}} \; := \;
\frac{8\pi G}{3 H_0^2c^2} \, \varepsilon_{\hbox{\scriptsize de}}^{\hbox{\scriptsize eff}}
\end{equation}
with the Hubble constant $H_0$ and the gravitational constant $G$.
This allows the comparison with the $\Lambda$CDM concordance model,
where $\Omega_{\hbox{\scriptsize de}}^{\hbox{\scriptsize eff}}$ of the UDM model
corresponds to the cold dark matter component and
$\Omega_{\hbox{\scriptsize de}}-\Omega_{\hbox{\scriptsize de}}^{\hbox{\scriptsize eff}}$
to the effective vacuum energy contribution.

The effective matter density $\Omega_{\hbox{\scriptsize de}}^{\hbox{\scriptsize eff}}$
can be computed for the generalised Chaplygin gas model leading to
\begin{equation}
\label{Def:Omega_de_eff_Chaplygin}
\Omega_{\hbox{\scriptsize Chap}}^{\hbox{\scriptsize eff}} \; = \;
\Omega_{\hbox{\scriptsize Chap}} \; (1-B)^{1/(1+\alpha)}
\hspace{10pt} \hbox{ for } \hspace{10pt}
\alpha > -1
\hspace{10pt} .
\end{equation}
This shows that
$\Omega_{\hbox{\scriptsize Chap}}-\Omega_{\hbox{\scriptsize Chap}}^{\hbox{\scriptsize eff}}$
becomes negative for $B<0$,
which would correspond in the $\Lambda$CDM case to a negative
$\Omega_\Lambda$ that is an anti-de Sitter like model.
We thus restrict the following analysis to $0<B<1$ and $\alpha>-1$.

In our analysis, the cosmological model is specified by the total
relative density
\begin{equation}
\label{Def:Omega_tot}
\Omega_{\hbox{\scriptsize tot}} \; = \; \Omega_{\hbox{\scriptsize rad}} \, + \,
\Omega_{\hbox{\scriptsize bar}} \, + \, \Omega_{\hbox{\scriptsize Chap}}
\hspace{10pt} .
\end{equation}
The radiation term $\Omega_{\hbox{\scriptsize rad}}$ takes the energy density
of photons and neutrinos with standard thermal history into account.
The baryon density $\Omega_{\hbox{\scriptsize bar}}$ is determined by the
choice of the Hubble constant $H_0 = 100\, h \hbox{ km}/(\hbox{s Mpc})$
as $\Omega_{\hbox{\scriptsize bar}} = 0.0223/h^2$,
so that the physical density of baryons is in agreement with the
Big-Bang nucleosynthesis,
i.\,e.\ according the measured deuterium to hydrogen abundance ratio
\cite{Pettini_Cooke_2012}.
As described below, the parameter estimation is carried out with the
Markov chains Monte Carlo (MCMC) algorithm and
in some selected cases, MCMC sequence are generated with an independently
varying $\Omega_{bar} h^2$ with $\Omega_{bar} h^2 = 0.0223\pm 0.0009$.
It is found that no change of the parameters defining the Chaplygin gas
cosmology $\alpha$ and $B$ occurs.
Since every additional parameter leads to longer MCMC sequences
until convergence is reached, we do not independently vary this parameter
for the further analysis in this paper.

Note, that in equation (\ref{Def:Omega_tot}) the dark sector is solely
described by the Chaplygin gas component which contrasts to other analyses
which add an extra cold dark matter term or a cosmological constant.
We have as model parameters the reduced Hubble constant $h$,
which in turn fixes $\Omega_{\hbox{\scriptsize bar}}$,
the density $\Omega_{\hbox{\scriptsize Chap}}$ together with the
parameters $\alpha$ and $B$, which specify the equation of state
(\ref{Eq:Chaplygin_equation_of_state}) of the Chaplygin gas.
The background dynamics is determined by
\begin{eqnarray}
\label{Eq:Hubble}
H(z) & = & H_0 \, \Big[
\Omega_{\hbox{\scriptsize rad}}(z+1)^4 +
\Omega_{\hbox{\scriptsize bar}}(z+1)^3 +
(1-\Omega_{\hbox{\scriptsize tot}})(z+1)^2
\\ & & \nonumber \hspace{70pt}  + \,
\Omega_{\hbox{\scriptsize Chap}} \, \{B + (1-B)(z+1)^{3(1+\alpha)}\}^{1/(1+\alpha)}
\Big]^{1/2}
\hspace{10pt} .
\end{eqnarray}

\section{The cosmological data sets}
\label{sec:data}

In this section, we describe the cosmological observations
that are used for the estimation of the model parameters
discussed in section \ref{sec:Chaplygin}.
The parameter estimation is carried out with the Markov chains Monte Carlo
(MCMC) algorithm which uses as parameters
$\alpha$ and $B$ occurring in the equation of state
(\ref{Eq:Def_eos_Chaplygin}) and (\ref{Eq:Chaplygin_energy_density}),
the density $\Omega_{\hbox{\scriptsize Chap}}$ and the reduced Hubble constant $h$.
In several cases, the parameter $\alpha$ is held fixed so that only
$B$, $\Omega_{\hbox{\scriptsize Chap}}$ and $h$ are varied.
The probability for selecting a new state in the Markov chain is
obtained from the $\chi^2_{\hbox{\scriptsize tot}}$ values
\begin{equation}
\label{Def:Chi2_total}
\chi^2_{\hbox{\scriptsize tot}} \; = \;
\chi^2_{\hbox{\scriptsize Sn}} \, + \, \chi^2_{\hbox{\scriptsize BAO}} \, + \, 
\chi^2_{\hbox{\scriptsize CMB}}
\hspace{10pt} ,
\end{equation}
whose individual contributions are discussed below.

In all cases the length of the MCMC sequences is at least 25\,000.
Several chains are generated and it is checked that they lead to the same
confidence contours.
Furthermore, the convergence of the Markov chains is checked by computing
the MCMC power spectrum and the convergence ratio $r$ along
the lines described in \cite{Dunkley_Bucher_Ferreira_Moodley_Skordis_2005}.

\subsection{The supernovae Ia data}

The supernovae Ia are now established standard candles
whose observed magnitude-redshift dependence can be compared
with the theoretical prediction of a given model.
We use the Union 2.1 compilation of the Supernova Cosmology Project
\cite{Suzuki_et_al_2012}
where the redshift and distance modulus $\mu_i^{\hbox{\scriptsize obs}}$
including its uncertainty $\sigma_{i,\hbox{\scriptsize Sn}}$
for $N_{\hbox{\scriptsize Sn}}=580$ supernovae Ia are given.
The calibration of the SNe Ia light curves makes use of a
fiducial $\Lambda$CDM model
so this data set possesses a model-dependence and
favours the $\Lambda$CDM cosmology.
From this Union 2.1 compilation, the $\chi^2$ value is computed
\begin{equation}
\label{Def:Chi2_Sn}
\chi^2_{\hbox{\scriptsize Sn}} \; = \;
\sum_{i=1}^{N_{\hbox{\scriptsize Sn}}} \,
\frac{(\mu_i^{\hbox{\scriptsize obs}}-\mu_i^{\hbox{\scriptsize th}}(z_i)-\Delta M)^2}
{\sigma_{i,\hbox{\scriptsize Sn}}^2}
\; + \;
\frac{\Delta M^2}{(0.15^{\hbox{\scriptsize mag}})^2}
\hspace{10pt} ,
\end{equation}
where the theoretical distance modulus $\mu_i^{\hbox{\scriptsize th}}(z_i)$
is computed from the luminosity distance $d_L(z)$ of the considered model.
The parameter $\Delta M$ allows a small variation of the
absolute magnitude $M_B\simeq -19.3^{\hbox{\scriptsize mag}}$ $(h=0.7)$
\cite{Suzuki_et_al_2012} of the supernovae Ia.
The parameter $\Delta M$ is analytically marginalised
\cite{Goliath_Amanullah_Astier_Goobar_Pain_2001}.
The order of the variation $0.15^{\hbox{\scriptsize mag}}$ is estimated according
\cite{Suzuki_et_al_2012}.
When the supernovae data are used without other data sets,
we constrain the variation of $\Delta M$ with the last summand
in (\ref{Def:Chi2_Sn}) because of the degeneracy
between the Hubble constant $h$ and the absolute magnitude $M_B$.
A change in the absolute magnitude from $M_B$ to $M_B'$ corresponds
to a shift of the Hubble constant $h$ to $h'$ with
\begin{equation}
\label{Def:Degeneracy_Magnitude_Hubble}
M_B \, - \, M_B' \; = \; 5 \log_{10}\left(\frac{h'}h\right)
\hspace{10pt} .
\end{equation}
The difference of the measurement of the Hubble constant of
\cite{Riess_et_al_2016}
with a best estimate of $H_0= 73.24\pm 1.74 \hbox{ km}/(\hbox{s Mpc})$
with the Planck value of $H_0= 67.8\pm 0.9 \hbox{ km}/(\hbox{s Mpc})$
\cite{Planck_2015_I} corresponds to a shift of the absolute magnitude of
$\Delta M \simeq 0.25^{\hbox{\scriptsize mag}}$.
When we generate Markov chains Monte Carlo together with other
data sets, we omit the last summand in (\ref{Def:Chi2_Sn})
and left $\Delta M$ unconstrained in the marginalisation.

\subsection{The BAO data}

While the supernovae data constrain primarily the luminosity distance $d_L(z)$,
the measurement of baryon acoustic oscillations allows the determination of
the angular diameter distance $d_A(z)$ as well as the value of
the Hubble length $d_H(z):=c/H(z)$.
The standard ruler for these lengths is provided by the sound horizon 
\begin{equation}
\label{Def:drag_sound_horizon}
r_s \; = \;
\int_{z_{\hbox{\scriptsize drag}}}^\infty \frac{c_s(z)}{H(z)}\, dz
\end{equation}
at the drag epoch at redshift $z_{\hbox{\scriptsize drag}}$
when photons and baryons decouple.
The speed of sound $c_s(z)$ in the photon-baryon fluid is given by
\begin{equation}
\label{Def:speed_of_sound}
c_s(z) \; = \;
\frac c{\sqrt{3(1+{\cal R}(z))}}
\hspace{10pt} \hbox{ with } \hspace{10pt}
{\cal R}(z) = \frac{3\rho_{\hbox{\scriptsize bar}}}{4\rho_\gamma} =
\frac 34 \frac{\Omega_{\hbox{\scriptsize bar}}}{\Omega_\gamma} \frac 1{z+1}
\hspace{10pt} .
\end{equation}
The drag redshift $z_{\hbox{\scriptsize drag}}$ can be determined by using
fitting formulae \cite{Eisenstein_Hu_1998,Aubourg_etc_2015}
but since we are dealing with cosmological models
that possibly leave the validity domain of the parameter space
where these fitting formulae are valid,
we use the exact approach.
A drag depth is defined as \cite{Hu_Sugiyama_1996}
\begin{equation}
\label{Def:drag_depth}
\tau_{\hbox{\scriptsize drag}}(z) \; := \;
\int_{\eta(z)}^{\eta_0} \frac{\dot\tau}{\cal R}\, d\eta
\hspace{10pt} ,
\end{equation}
where $\tau$ is the Compton optical depth with $\dot\tau$ denoting its
derivative with respect to conformal time $\eta$ and $d\eta = c\,dt/a(t)$.
The drag epoch is then determined by the redshift $z_{\hbox{\scriptsize drag}}$
satisfying \cite{Hu_Sugiyama_1996}
\begin{equation}
\label{Def:drag_epoch}
\tau_{\hbox{\scriptsize drag}}(z_{\hbox{\scriptsize drag}}) = 1
\hspace{10pt} .
\end{equation}
With $z_{\hbox{\scriptsize drag}}$ computed in this way,
the equation (\ref{Def:drag_sound_horizon}) gives the standard ruler $r_s$
which in turn allows the conversion of the BAO data
such that a $\chi^2$ analysis of the Chaplygin gas cosmology is possible.

In table \ref{Tab:BAO_data} the BAO data used in this paper are listed.
For redshifts $z<0.2$ the data of the 6dF galaxy survey
\cite{Beutler_et_al_2011} and the SDSS Data Release 7 main galaxy sample
\cite{Ross_et_al_2015} are used
which are given in terms of volume averaged distances
$$
d_V(z) \; := \; \sqrt[3]{z\, (1+z)^2 \, d_H(z) d_A^2(z)} 
\hspace{10pt} .
$$
For $0.2<z<1$, the Data Release 12 of the SDSS-III/BOSS spectroscopic galaxy
sample \cite{Chuang_et_al_2016} is used,
which are also given in table \ref{Tab:BAO_data}.
The data for $H(z)$ and $d_A(z)$ are used simultaneously.
Furthermore, the BAO data extracted from the auto-correlation of the
Lyman $\alpha$ forest fluctuations of the BOSS Data Release 11 quasars
\cite{Delubac_et_al_2015} are used.
In addition, the BAO data obtained from the quasar-Lyman $\alpha$
cross-correlation of the Data Release 11 of the SDSS-III/BOSS
\cite{Font_Ribera_et_al_2014} are taken into account.
Although both Lyman $\alpha$ data sets are derived from the same volume,
they can be considered as independent data points as stated in
\cite{Aubourg_etc_2015}.
The auto-correlation and cross-correlation approaches are complementary
because of the quite different impact of redshift-space distortion on the
two measurements.
These BAO data can be considered as independent
because their uncertainties are not dominated by cosmic variance,
but instead are dominated by the combination of noise in the spectra and
sparse sampling of the structure in the survey volume,
both of which affect the auto-correlation and cross-correlation almost
independently \cite{Aubourg_etc_2015}.
In order to substantiate this, 
tests with mock catalogues and several analysis procedures are carried out in
\cite{Delubac_et_al_2015},
which find a good agreement between error estimates from the likelihood function
and from the variance in mock catalogues.
Thus, in our $\chi^2$ analysis the $d_A(z)$ and $d_H(z)$ values of both
Lyman $\alpha$ data sets are taken into account.
It is worthwhile to note that the BAO data are biased towards the
$\Lambda$CDM model especially for larger values of the redshift $z$,
because of the necessary distance calibrations.
Thus, the BAO data possess a model-dependence in favour of the concordance
model.

The $\chi^2$ values are computed in the usual way as
\begin{equation}
\label{Def:Chi2_BAO}
\chi^2_{\hbox{\scriptsize BAO}} \; = \;
\sum_{i=1}^{N_{\hbox{\scriptsize BAO}}} \,
\frac{(X_i^{\hbox{\scriptsize obs}}-X_i^{\hbox{\scriptsize th}}(z_i))^2}
{\sigma_{i,\hbox{\scriptsize BAO}}^2}
\hspace{10pt} ,
\end{equation}
where $X_i$ stands for one of the corresponding distances
listed in table \ref{Tab:BAO_data}.

\begin{table}[ph]
\hspace{-15pt}\hspace{-25pt}\begin{tabular}{|c|c|c|c|c|c|}
\hline
redshift $z$ & $H(z) r_s$ & $d_A(z)/{r_s}$ & $d_V(z)/{r_s}$ &
$d_H(z)/{r_s}$  & Reference \\
\hline
0.106  & & &  $3.06\pm 0.13$  &  & 6dFGS \cite{Beutler_et_al_2011} \\
0.15   & & &  $4.47\pm 0.17$  &  & SDSS DR7 \cite{Ross_et_al_2015} \\
\hline
0.24 &  $11636.0\pm 827$ &  $5.59\pm 0.30$ & & & \\
0.32 &  $11075.0\pm 591$ &  $6.47\pm 0.19$ & & & \\
0.37 &  $11045.0\pm 930$ &  $6.72\pm 0.44$ & & &
SDSS-III/BOSS \\
0.49 &  $12920.0\pm 709$ &  $8.72\pm 0.21$ & & &
DR12 \cite{Chuang_et_al_2016} \\
0.59 &  $14279.0\pm 399$ &  $9.62\pm 0.16$ & & & \\
0.64 &  $14530.0\pm 546$ &  $9.78\pm 0.28$ & & & \\
\hline
2.34 & & $11.28\pm 0.65$ & & $9.18\pm 0.28$ &
DR11 Ly$\alpha$ (auto) \cite{Delubac_et_al_2015} \\
2.36 & & $10.8\pm 0.4$   & & $9.00\pm 0.30$ &
DR11 Ly$\alpha$ (cross) \cite{Font_Ribera_et_al_2014} \\
\hline
\end{tabular}
\caption{\label{Tab:BAO_data}
The BAO data used for constraining the generalised Chaplygin gas cosmology.
}
\end{table}

\subsubsection{On the significance of the Lyman $\alpha$ data}
\label{sec:Lya}

\begin{figure}[ph]
\begin{center}
\hspace*{-50pt}\begin{minipage}{20cm}
\vspace*{-10pt}
\hspace*{-20pt}\includegraphics[width=9.0cm]{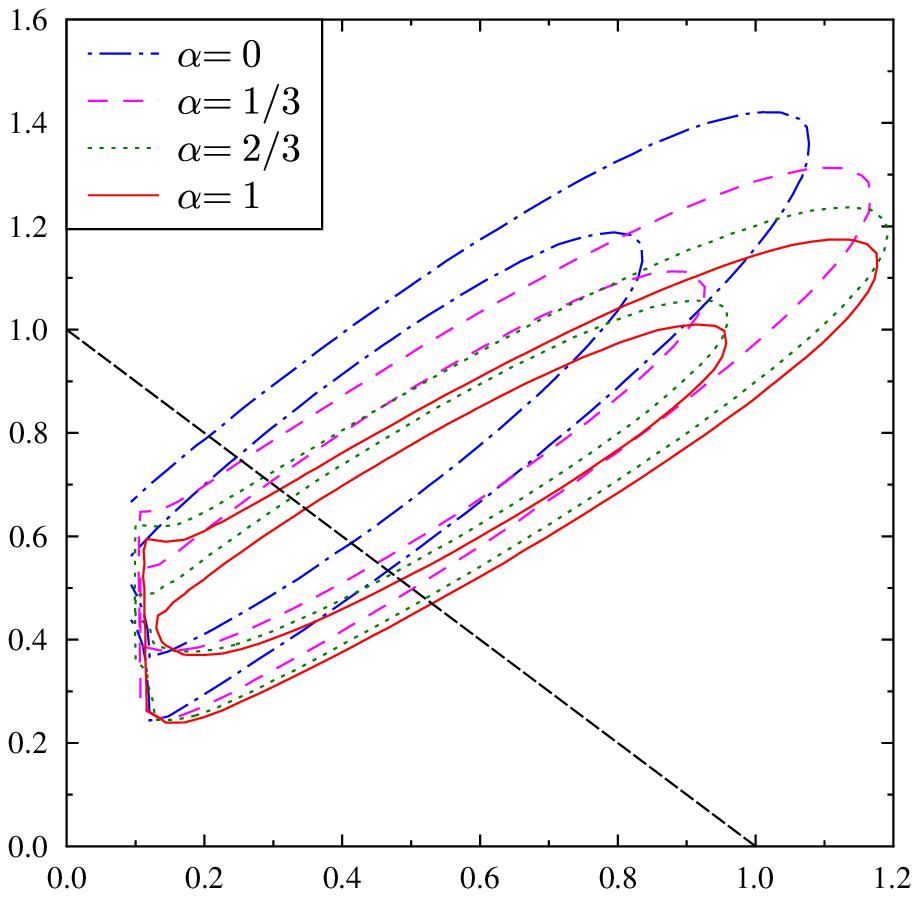}
\hspace*{-20pt}\includegraphics[width=9.0cm]{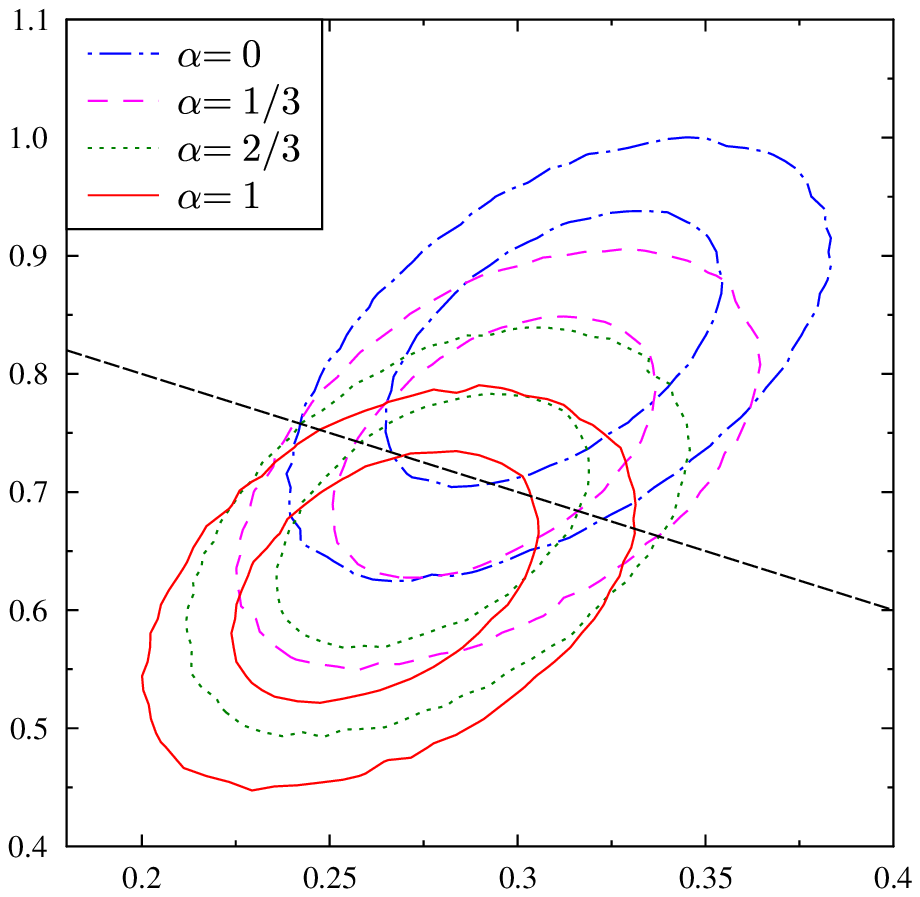}
\put(-380,195){(a)}
\put(-140,195){(b)}
\put(-160,14){$\Omega_{\hbox{\scriptsize bar}}+\Omega_{\hbox{\scriptsize Chap}}^{\hbox{\scriptsize eff}}$}
\put(-250,95){\rotatebox{90}{$\Omega_{\hbox{\scriptsize Chap}}-\Omega_{\hbox{\scriptsize Chap}}^{\hbox{\scriptsize eff}}$}}
\put(-400,14){$\Omega_{\hbox{\scriptsize bar}}+\Omega_{\hbox{\scriptsize Chap}}^{\hbox{\scriptsize eff}}$}
\put(-490,95){\rotatebox{90}{$\Omega_{\hbox{\scriptsize Chap}}-\Omega_{\hbox{\scriptsize Chap}}^{\hbox{\scriptsize eff}}$}}
\end{minipage}
\vspace*{-25pt}
\end{center}
\caption{\label{Fig:bao_alpha_seq_noLya}
The 1$\sigma$ and 2$\sigma$ confidence contours are shown with respect to
effective matter and energy density for four Chaplygin models with
fixed values of $\alpha$.
The panel (a) uses the BAO data without the Ly$\alpha$ values,
whereas the panel (b) includes them.
Note the different scales on the axes.
The dashed line corresponds to universes with $\Omega_{\hbox{\scriptsize tot}}=1$.
}
\end{figure}

\begin{figure}[ph]
\begin{center}
\hspace*{-50pt}\begin{minipage}{20cm}
\vspace*{-10pt}
\hspace*{-20pt}\includegraphics[width=9.0cm]{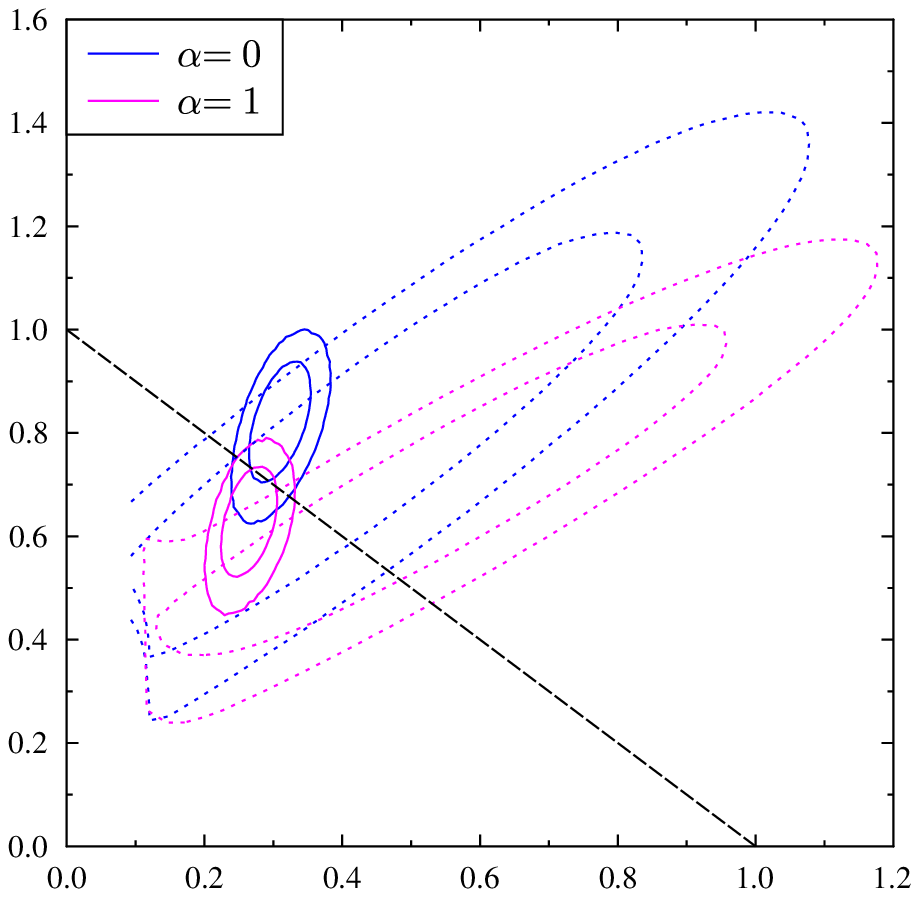}
\hspace*{-20pt}\includegraphics[width=9.0cm]{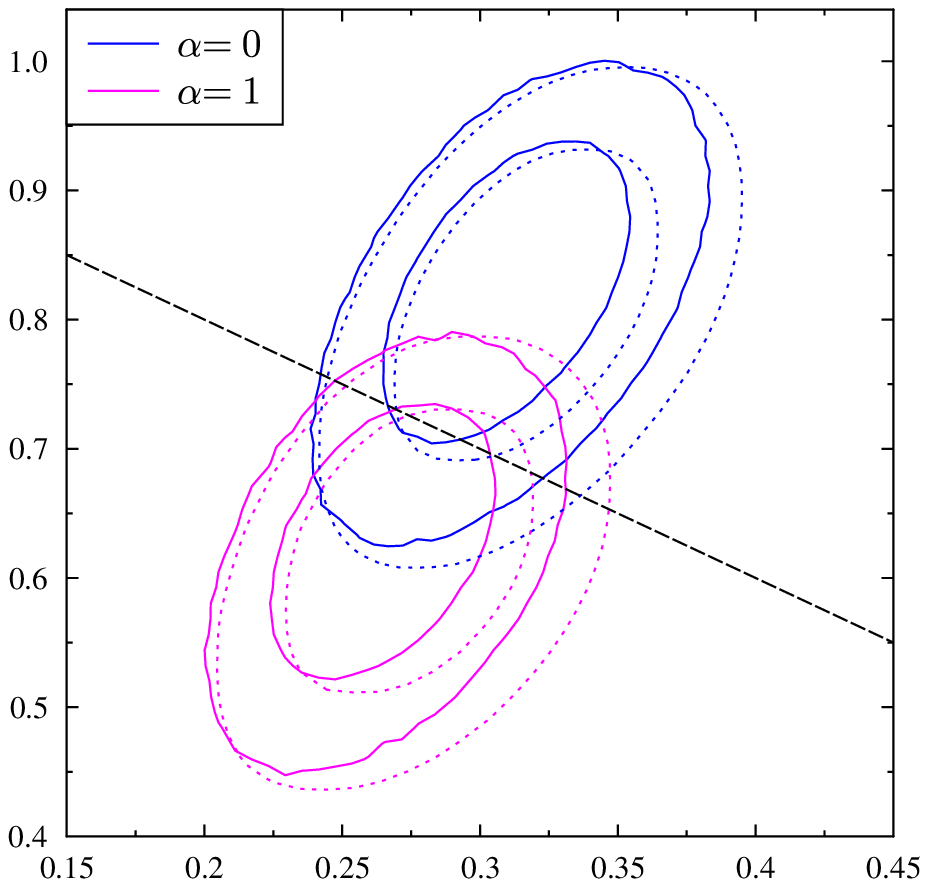}
\put(-395,195){(a)}
\put(-155,195){(b)}
\put(-160,14){$\Omega_{\hbox{\scriptsize bar}}+\Omega_{\hbox{\scriptsize Chap}}^{\hbox{\scriptsize eff}}$}
\put(-250,95){\rotatebox{90}{$\Omega_{\hbox{\scriptsize Chap}}-\Omega_{\hbox{\scriptsize Chap}}^{\hbox{\scriptsize eff}}$}}
\put(-400,14){$\Omega_{\hbox{\scriptsize bar}}+\Omega_{\hbox{\scriptsize Chap}}^{\hbox{\scriptsize eff}}$}
\put(-490,95){\rotatebox{90}{$\Omega_{\hbox{\scriptsize Chap}}-\Omega_{\hbox{\scriptsize Chap}}^{\hbox{\scriptsize eff}}$}}
\end{minipage}
\vspace*{-25pt}
\end{center}
\caption{\label{Fig:bao_alpha_vgl_Lya}
To emphasise the restricting power of the Ly$\alpha$ data around $z=2.34$,
the 1$\sigma$ and 2$\sigma$ confidence contours are shown with respect to
effective matter and effective dark energy density for two Chaplygin models
with $\alpha=0$ and $\alpha=1$.
The dotted curves in panel (a) are computed using only the BAO data below
$z=1$ as in figure \ref{Fig:bao_alpha_seq_noLya}(a),
while the full curves take also the Lyman $\alpha$ forest derived data of
table \ref{Tab:BAO_data} into account.
The panel (b) shows the shift of the confidence domains
when the auto-correlation derived Lyman $\alpha$ DR11 data (full curves)
are replaced with those of DR12 (dotted curves)
while leaving the other BAO data unchanged.
The dashed line indicates the positions of the models with
$\Omega_{\hbox{\scriptsize tot}}=1$ in both panels.
}
\end{figure}

The addition of the Lyman $\alpha$ data to the $\chi^2$ analysis
significantly constrains the models of the Chaplygin cosmology
due to their larger redshift.
In order to show this, the figures \ref{Fig:bao_alpha_seq_noLya} and
\ref{Fig:bao_alpha_vgl_Lya} show the 1$\sigma$ and 2$\sigma$ confidence domains
for several Chaplygin cosmologies with a fixed value of $\alpha$
as defined in equation (\ref{Eq:Def_eos_Chaplygin}) using the BAO data
of table \ref{Tab:BAO_data} with and without the Lyman $\alpha$ data.
The figures do not show the confidence domains with respect to the
parameters that are varied in the MCMC algorithm,
but instead, they show the combinations
$\Omega_{\hbox{\scriptsize Chap}}-\Omega_{\hbox{\scriptsize Chap}}^{\hbox{\scriptsize eff}}$ and
$\Omega_{\hbox{\scriptsize bar}}+\Omega_{\hbox{\scriptsize Chap}}^{\hbox{\scriptsize eff}}$
where $\Omega_{\hbox{\scriptsize Chap}}^{\hbox{\scriptsize eff}}$ is defined in
(\ref{Def:Omega_de_eff_Chaplygin}).
For the special case $\alpha=0$ corresponding to the $\Lambda$CDM cosmology,
the combination 
$\Omega_{\hbox{\scriptsize Chap}}-\Omega_{\hbox{\scriptsize Chap}}^{\hbox{\scriptsize eff}}$
corresponds to the contribution of the cosmological constant $\Omega_\Lambda$ and
$\Omega_{\hbox{\scriptsize bar}}+\Omega_{\hbox{\scriptsize Chap}}^{\hbox{\scriptsize eff}}$
to the matter content $\Omega_{\hbox{\scriptsize mat}}$
as outlined in section \ref{sec:Chaplygin}.

The figure \ref{Fig:bao_alpha_seq_noLya}(a) shows the confidence domains
without the constraining power of the Lyman $\alpha$ data,
that is using the values of table \ref{Tab:BAO_data} with $z<1$,
for four Chaplygin cosmologies with $\alpha=0$, $\frac 13$, $\frac 23$, and 1.
The figure reveals very large confidence domains.
The cut-off like behaviour of the confidence domains around
$\Omega_{\hbox{\scriptsize bar}}+\Omega_{\hbox{\scriptsize Chap}}^{\hbox{\scriptsize eff}}
\simeq 0.16$ is due to the restriction $B<1$ which avoids
phantom-like energy at the current epoch and a singular behaviour of
the equation of state in the past (see section \ref{sec:Chaplygin}).

Including the Lyman $\alpha$ data reduces drastically the confidence domains
so that the restriction to $B<1$ does not become appreciable as shown
in figure \ref{Fig:bao_alpha_seq_noLya}(b).
It is seen that with increasing values of $\alpha$,
the best-fit region prefers lower densities of $\Omega_{\hbox{\scriptsize tot}}$.
While the Chaplygin cosmology with $\alpha=0$ prefers models with a
positive spatial curvature ($\Omega_{\hbox{\scriptsize tot}}>1$),
the revers is true for the case $\alpha=1$.
The figure \ref{Fig:bao_alpha_vgl_Lya}(a) displays 
the 1$\sigma$ and 2$\sigma$ confidence domains for $\alpha=0$ and
for $\alpha=1$ with and without the Lyman $\alpha$ data in one plot and
emphasises the restrictions due to them.

Recently, the SDSS DR12 data have been used to compute the
auto-correlation of the Lyman $\alpha$ forest fluctuations and
to derive updated values of $d_A(z)$ and $d_H(z)$ for $z=2.33$
\cite{Bautista_et_al_2017}.
The updated distances
$d_A(z=2.33)/r_s=11.34\pm 0.64$ and $d_H(z=2.33)/r_s=9.07\pm 0.31$
agree with the previous Lyman $\alpha$ results derived from
the Data Release 11 \cite{Delubac_et_al_2015,Font_Ribera_et_al_2014}
within $\sim 0.5\,\sigma$ \cite{Bautista_et_al_2017}.
In order to compare these updated auto-correlation based new data,
two Monte Carlo chains are generated,
where one chain uses the BAO data as listed in table \ref{Tab:BAO_data}
with the Lyman $\alpha$ derived values of DR11.
In the other chain, the auto-correlation derived DR11 values of
\cite{Delubac_et_al_2015} are replaced by the updated DR12 values of
\cite{Bautista_et_al_2017}.
The result is shown in figure \ref{Fig:bao_alpha_vgl_Lya}(b).
The confidence regions are shifted relative to each other,
but there is nevertheless a large overlap between them.
This demonstrates the sensitivity due to the data point around $z=2.34$.
However, as it is also the case for the supernovae data,
the BAO data and especially the Lyman $\alpha$ forest data are
also contaminated by the $\Lambda$CDM model
which enters in the distance calibration.

\subsection{The cosmic microwave background data of Planck 2015}
\label{sec:CMB}

The supernovae Ia and BAO data discussed above provide
likelihood estimations for cosmological models with respect to their
background behaviour.
The angular power spectrum
${\cal D}_l^{\hbox{\scriptsize TT}} := l(l+1) C_l^{\hbox{\scriptsize TT}}/(2\pi)$
of the cosmic microwave background radiation,
on the other hand, tests the evolution of perturbations.
The multipole spectrum $C_l^{\hbox{\scriptsize TT}}$ is defined by
$$
C_l^{\hbox{\scriptsize TT}} \; = \; \frac 1{2l+1} \sum_{m=-l}^l |a_{lm}|^2
$$
where $a_{lm}$ are the expansion coefficients of the temperature fluctuations
with respect to the spherical harmonics.
In this paper, the ${\cal D}_l^{\hbox{\scriptsize TT}}$ measurements of the
Planck 2015 data release \cite{Planck_2015_I} are used.
The likelihood for a theoretical angular power spectrum
${\cal D}_l^{\hbox{\scriptsize TT}}$ is calculated with the 
Planck Likelihood Code R2.00 provided on the Planck website
\texttt{http://pla.esac.esa.int/pla/\#cosmology}
which computes the log likelihoods of the temperature maps.
These log likelihoods are then converted to the corresponding
$\chi^2_{\hbox{\scriptsize CMB}}$ to be used in (\ref{Def:Chi2_total}).
The low-$l$ likelihood based on the results of the Commander approach
is calculated for $l=2$ to 29 using the file
\texttt{commander\_rc2\_v1.1\_l2\_29\_B.clik}
which is also given at the Planck website.
Since the ${\cal D}_l^{\hbox{\scriptsize TT}}$ spectra for the
Chaplygin gas cosmology are computed up to $l=1000$,
we use for the high-$l$ likelihood
the file \texttt{plik\_dx11dr2\_HM\_v18\_TT.clik} ($29<l<2509$)
modified to cover the multipole range $29<l\leq 1000$.
The restriction to the multipole range $l\leq 1000$
reduces significantly the CPU time of the MCMC estimations.

The ${\cal D}_l^{\hbox{\scriptsize TT}}$ spectra for the Chaplygin gas cosmology
are computed with the code described in \cite{Aurich_Lustig_2016}.
In this code, the isentropic initial conditions are specified at
conformal time $\eta=0$.
Then, power series in $\eta$ for the various perturbations are used to
analytically compute the perturbations at $\eta=\eta_{\hbox{\scriptsize rec}}/200$,
where $\eta_{\hbox{\scriptsize rec}}$ denotes the time of recombination.
With these values, the Boltzmann equations are integrated numerically
up to the present epoch leading to the theoretical
${\cal D}_l^{\hbox{\scriptsize TT}}$ spectra.
The perturbation power series, which are derived in \cite{Aurich_Lustig_2016},
allow a faithful treatment of the initial conditions towards a later time,
since for $\eta\to 0$, a numerical integration of the system of differential
equations is not possible due to the singular behaviour of various terms
at $\eta=0$.
This singular behaviour leads to a cancellation of numerical digits by
the subtraction of very large terms of the same magnitude.
This difficulty is circumvented by using analytic power series in the
interval $\eta\in[0,\eta_{\hbox{\scriptsize rec}}/200]$.
Since there are no such accuracy problems
for $\eta > \eta_{\hbox{\scriptsize rec}}/200$,
the integration of the perturbation equations can then be done numerically
for these large values of $\eta$.
The validity of our code is checked by comparing the
${\cal D}_l^{\hbox{\scriptsize TT}}$ spectra for the wCDM model
for a wide range of curvatures obtained
from our code with those computed with the public software
CAMB\footnote{The software is available at http:/$\!$/camb.info}.

Besides the equation of state, there are further degrees of freedom
which characterise the Chaplygin gas, or a general dark matter component,
as emphasised by Hu \cite{Hu_1998},
where a generalised dark matter component is introduced
which is defined by the effective velocity of sound $c_{\hbox{\scriptsize eff}}$
(in the rest frame of the dark component) and a viscosity velocity
$c_{\hbox{\scriptsize vis}}$ which is related to the anisotropic stress.
The effective speed of sound can be interpreted as
the rest frame speed of sound
\begin{equation}
c^2_{\hbox{\scriptsize eff}} \; = \;
\frac{\delta p_g^{\hbox{\scriptsize (rest)}}}
{\delta \rho_g^{\hbox{\scriptsize (rest)}}}
\hspace{10pt} ,
\end{equation}
where the pressure and density perturbations in the rest frame
of the generalised dark matter component are denoted by
$\delta p_g^{\hbox{\scriptsize (rest)}}$ and $\delta \rho_g^{\hbox{\scriptsize (rest)}}$.
In this paper, both velocities are set to zero
$c^2_{\hbox{\scriptsize eff}} = c^2_{\hbox{\scriptsize vis}} = 0$.
For a UDM model, ${\cal D}_l^{\hbox{\scriptsize TT}}$ spectra for
a positive $c^2_{\hbox{\scriptsize eff}}$ are shown in
figures 7 and 8 in \cite{Aurich_Lustig_2016}.

The theoretical ${\cal D}_l^{\hbox{\scriptsize TT}}$ spectra depend on
the scalar spectral index $n_s$ and the overall normalisation.
After computing the CMB transfer function for a model of the MCMC sequence,
these two quantities are determined such that the value of
$\chi^2_{\hbox{\scriptsize CMB}}$ computed by the Planck Likelihood Code is minimised.
Thus, neither $n_s$ nor the overall normalisation occur as free parameters
in the MCMC sequence.
Furthermore, the reionisation is modelled by a smooth transition with
a width $\Delta z=0.4$ at a redshift $z_{\hbox{\scriptsize reion}}=10.6$,
at which half of the matter is reionised.
The choice of $z_{\hbox{\scriptsize reion}}$ and $\Delta z$ does not sensitively
influence the likelihood, see e.\,g.\ figure 6 in \cite{Aurich_Lustig_2016},
where ${\cal D}_l^{\hbox{\scriptsize TT}}$ is computed for two different values
of the reionisation optical depth.
The ${\cal D}_l^{\hbox{\scriptsize TT}}$ spectra are only changed at small values
of $l$, where the data possess large uncertainties.

\section{Analysis and Results}
\label{sec:Analysis_and_Results}

\subsection{MCMC analysis for a fixed value of $\alpha$}
\label{sec:MCMC_with_fixed_alpha}

In this subsection, the best-fit estimates are considered
for fixed values of the equation of state parameter $\alpha$.
The aim is to reveal which data set is responsible for the shift
of the best-fit parameters.
On the one hand, MCMC sequences are generated
which take into account only one of the data sets BAO, SN and CMB
as discussed above.
Their $\chi^2$ values are denoted as $\chi^2_{\hbox{\scriptsize BAO}}$,
$\chi^2_{\hbox{\scriptsize SN}}$ and $\chi^2_{\hbox{\scriptsize CMB}}$
and the best-fit models minimise only one of the three $\chi^2$ values
and ignore the other two.
On the other hand, further MCMC sequences are generated that take all
three data sets simultaneously into account
whose $\chi^2$ value is the corresponding sum
\begin{equation}
\label{Def:chi2_total}
\chi^2_{\hbox{\scriptsize BAO+SN+CMB}} \; = \;
\chi^2_{\hbox{\scriptsize BAO}} \, + \, \chi^2_{\hbox{\scriptsize SN}} \, + \,
\chi^2_{\hbox{\scriptsize CMB}}
\end{equation}
and their best-fit model minimises $\chi^2_{\hbox{\scriptsize BAO+SN+CMB}}$.
Thus, we consider four kinds of best-fit models which minimise
$\chi^2_{\hbox{\scriptsize BAO}}$, $\chi^2_{\hbox{\scriptsize SN}}$,
$\chi^2_{\hbox{\scriptsize CMB}}$, and $\chi^2_{\hbox{\scriptsize BAO+SN+CMB}}$, respectively.
The individual components in (\ref{Def:chi2_total}) belonging to the smallest
total value $\chi^2_{\hbox{\scriptsize BAO+SN+CMB}}$ are denoted as
$\bar\chi^2_{\hbox{\scriptsize BAO}}$, $\bar\chi^2_{\hbox{\scriptsize SN}}$, and
$\bar\chi^2_{\hbox{\scriptsize CMB}}$.
These are, in general, larger than the smallest $\chi^2$ values of the
MCMC sequences which take only one data set into account.
The difference between the smallest values of
$\chi^2_{\hbox{\scriptsize BAO}}$, $\chi^2_{\hbox{\scriptsize SN}}$, and
$\chi^2_{\hbox{\scriptsize CMB}}$ with the corresponding values of
$\bar\chi^2_{\hbox{\scriptsize BAO}}$, $\bar\chi^2_{\hbox{\scriptsize SN}}$, and
$\bar\chi^2_{\hbox{\scriptsize CMB}}$ reveals the tension between
the three data sets with respect to a cosmological model.
This difference indicates how far the optimal cosmological parameters with
respect to a single data set are driven away in order to yield the best
compromise with respect to all three data sets.
In the case of a perfect concordance between the three data sets,
one would obtain $\chi^2_{\hbox{\scriptsize BAO}}=\bar\chi^2_{\hbox{\scriptsize BAO}}$,
$\chi^2_{\hbox{\scriptsize SN}}=\bar\chi^2_{\hbox{\scriptsize SN}}$, and
$\chi^2_{\hbox{\scriptsize CMB}}=\bar\chi^2_{\hbox{\scriptsize CMB}}$.


\begin{table}[ph]
\begin{tabular}{|c|c|c|c|c|c|c|c|}
\hline
$\alpha$ & $\chi^2_{\hbox{\scriptsize BAO}}$ &  $\bar\chi^2_{\hbox{\scriptsize BAO}}$ & 
$\chi^2_{\hbox{\scriptsize SN}}$ &  $\bar\chi^2_{\hbox{\scriptsize SN}}$ & 
$\chi^2_{\hbox{\scriptsize CMB}}$ &  $\bar\chi^2_{\hbox{\scriptsize CMB}}$ & 
$\chi^2_{\hbox{\scriptsize BAO+SN+CMB}}$ \\
\hline
-2/3  & 11.94 & 96.49 & 562.34 & 764.92 & 1376.69 & 1389.37 & 2250.78 \\
-1/3  & 12.30 & 33.35 & 562.26 & 602.37 & 1385.99 & 1392.28 & 2028.00 \\
\hline
0   & 18.46 & 21.33 & 562.23 & 563.05 & 1390.91 & 1393.43 & 1977.80 \\
1/3 & 25.81 & 26.61 & 562.20 & 571.21 & 1394.71 & 1395.67 & 1993.49 \\
2/3 & 32.04 & 36.44 & 562.18 & 594.57 & 1397.13 & 1398.45 & 2029.45 \\
1   & 36.74 & 47.26 & 562.16 & 620.07 & 1398.90 & 1400.95 & 2068.28 \\
\hline
\end{tabular}
\caption{\label{Tab:Chi2_data}
The minimum of $\chi^2$ is given for several fixed values of the
equation of state parameter $\alpha$.
The columns with $\chi^2_{\hbox{\scriptsize BAO}}$, $\chi^2_{\hbox{\scriptsize SN}}$
and $\chi^2_{\hbox{\scriptsize CMB}}$ list the minimum of $\chi^2$,
if only the corresponding data set is used.
The column with $\chi^2_{\hbox{\scriptsize BAO+SN+CMB}}$ lists the minimum
if all three data sets are used simultaneously, and
$\bar\chi^2_{\hbox{\scriptsize BAO}}$, $\bar\chi^2_{\hbox{\scriptsize SN}}$,
$\bar\chi^2_{\hbox{\scriptsize CMB}}$ are the individual components contributing to
the minimal value of $\chi^2_{\hbox{\scriptsize BAO+SN+CMB}}$.
}
\end{table}



\begin{table}[ph]
\begin{tabular}{|c|c|c|c|c|c|c|c|}
\hline
$\alpha$ & $\chi^2_{\hbox{\scriptsize BAO}}$ & $\bar\chi^2_{\hbox{\scriptsize BAO}}$ & 
$\chi^2_{\hbox{\scriptsize SN}}$ &  $\bar\chi^2_{\hbox{\scriptsize SN}}$ & 
$\chi^2_{\hbox{\scriptsize CMB}}$ &  $\bar\chi^2_{\hbox{\scriptsize CMB}}$ & 
$\chi^2_{\hbox{\scriptsize BAO+SN+CMB}}$ \\
\hline
0   & 9.62 & 10.52 & 562.23 & 563.03 & 1390.91 & 1393.28 & 1966.83 \\
1/3 & 9.17 & 15.48 & 562.20 & 569.52 & 1394.71 & 1396.23 & 1981.23 \\
2/3 & 8.84 & 23.79 & 562.18 & 591.77 & 1397.13 & 1399.11 & 2014.67 \\
1   & 8.66 & 31.11 & 562.16 & 618.36 & 1398.90 & 1401.74 & 2051.20 \\
\hline
\end{tabular}
\caption{\label{Tab:Chi2_data_without_Lya}
In contrast to table \ref{Tab:Chi2_data}, we use here only the BAO data
without those derived from the Lyman $\alpha$ forest,
that is without the BAO values at $z\simeq 2.34$.
All other data sets are the same as used in table \ref{Tab:Chi2_data}.
}
\end{table}


In table \ref{Tab:Chi2_data} the minimal values of $\chi^2$ are given
for the generalised Chaplygin gas model for several fixed values of the
parameter $\alpha$ as defined in (\ref{Eq:Def_eos_Chaplygin}).
The table reveals that the supernovae data in isolation are
the least stringent ones
since they always give $\chi^2_{\hbox{\scriptsize SN}}$ values around 562.2
independently of $\alpha$.
At second comes the CMB data, whose $\chi^2_{\hbox{\scriptsize CMB}}$ values
increases by 8 from $\alpha=0$ to $\alpha=1$.
In contrast, the BAO data lead for the same range of $\alpha$ to the largest
increase by 18.3.
Concerning the values of $\alpha$ given in table \ref{Tab:Chi2_data},
one concludes that the data sets prefer the value
$\alpha=0$ which corresponds to the $\Lambda$CDM model.
It is interesting to note that the CMB data alone possesses even smaller
values of $\chi^2_{\hbox{\scriptsize CMB}}$ for negative values of $\alpha$
as revealed by table \ref{Tab:Chi2_data}.

\begin{figure}
\begin{center}
\hspace*{-50pt}\begin{minipage}{20cm}
\begin{minipage}{20cm}
\vspace*{-30pt}
\hspace*{-40pt}\includegraphics[width=10.0cm]{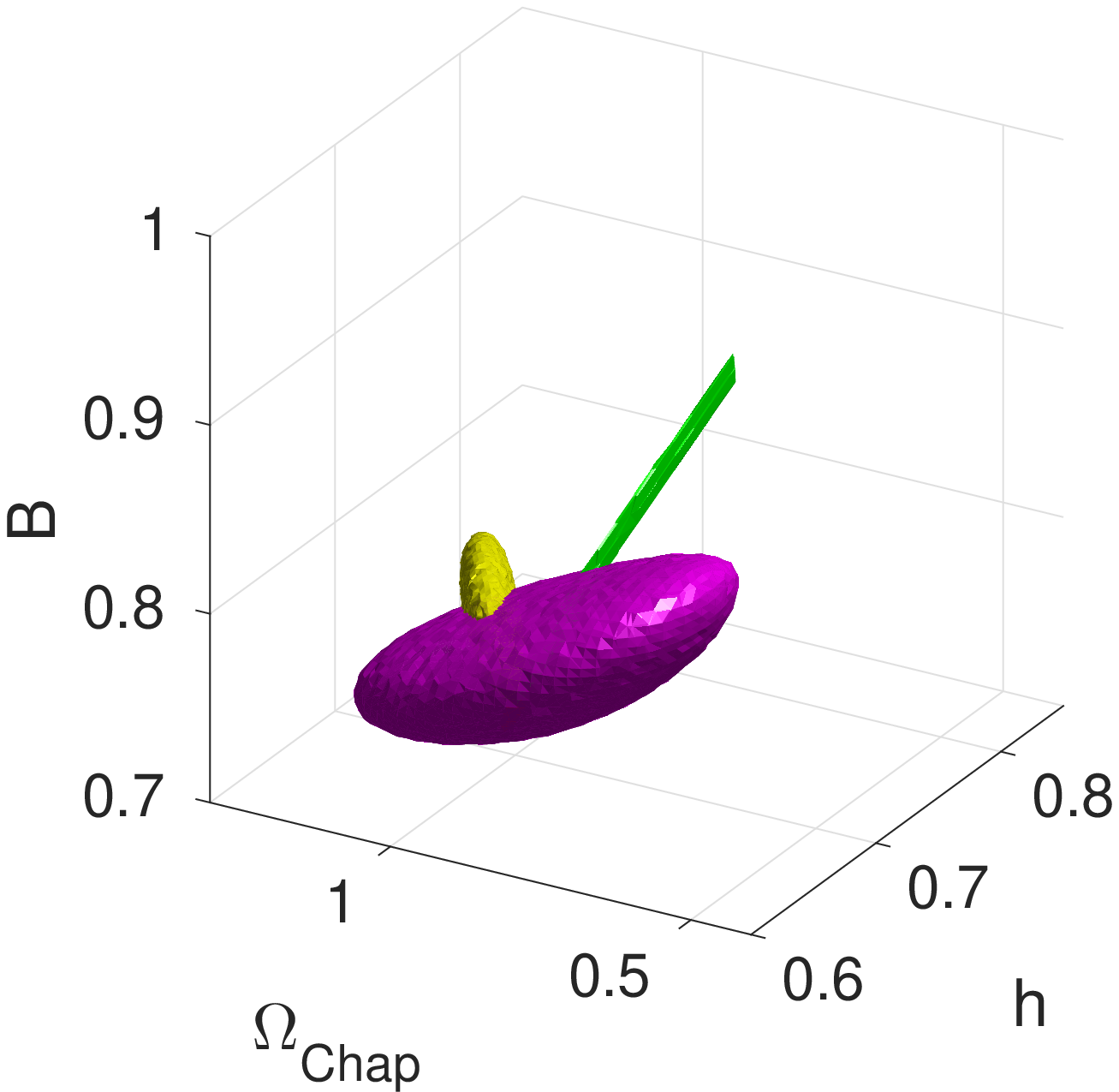}
\hspace*{-40pt}\includegraphics[width=10.0cm]{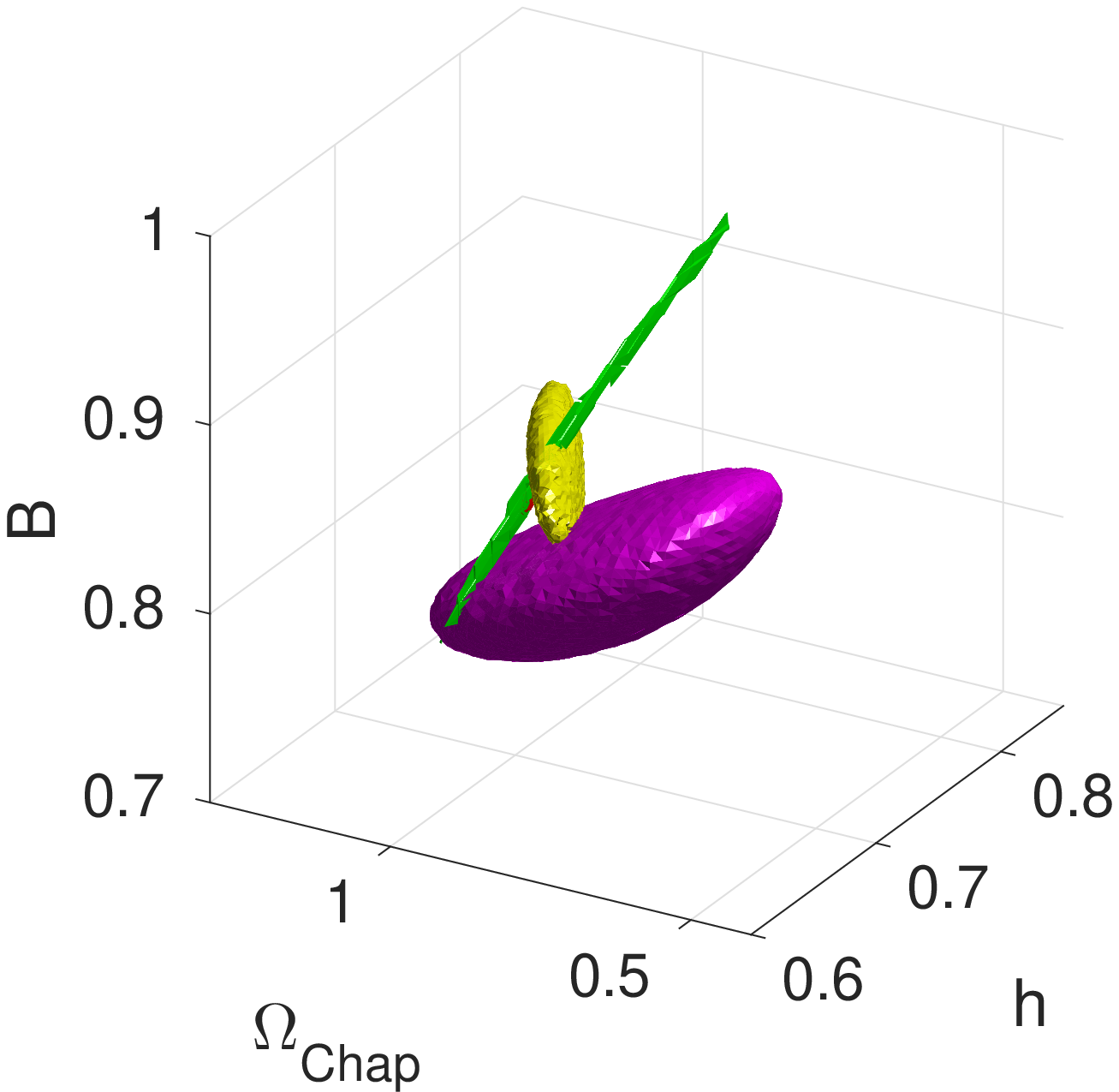}
\put(-490,170){(a) \hspace{20pt} $\alpha = 0$}
\put(-220,170){(b) \hspace{20pt} $\alpha = 1/3$}
\end{minipage}
\begin{minipage}{20cm}
\hspace*{-40pt}\includegraphics[width=10.0cm]{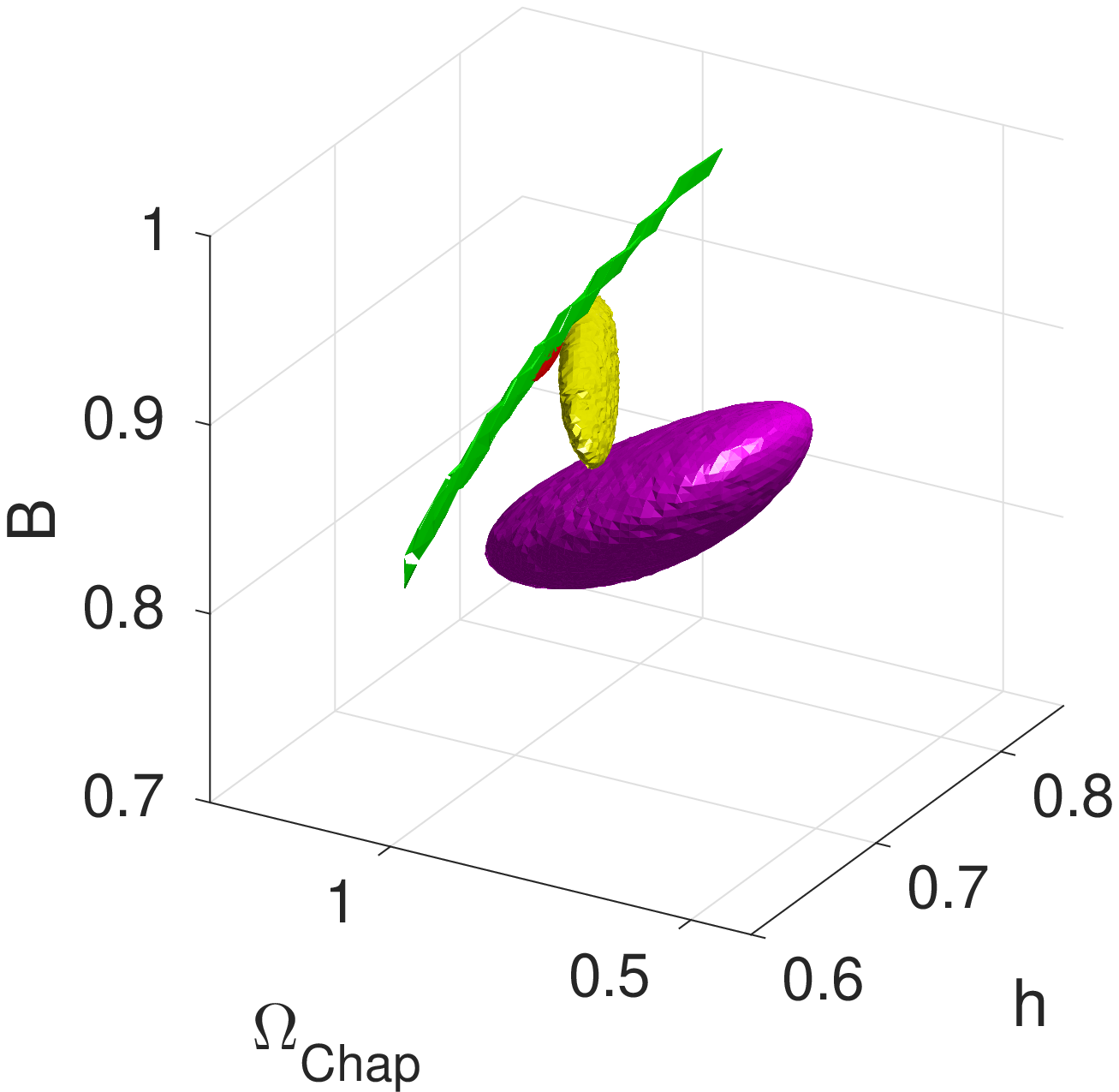}
\hspace*{-40pt}\includegraphics[width=10.0cm]{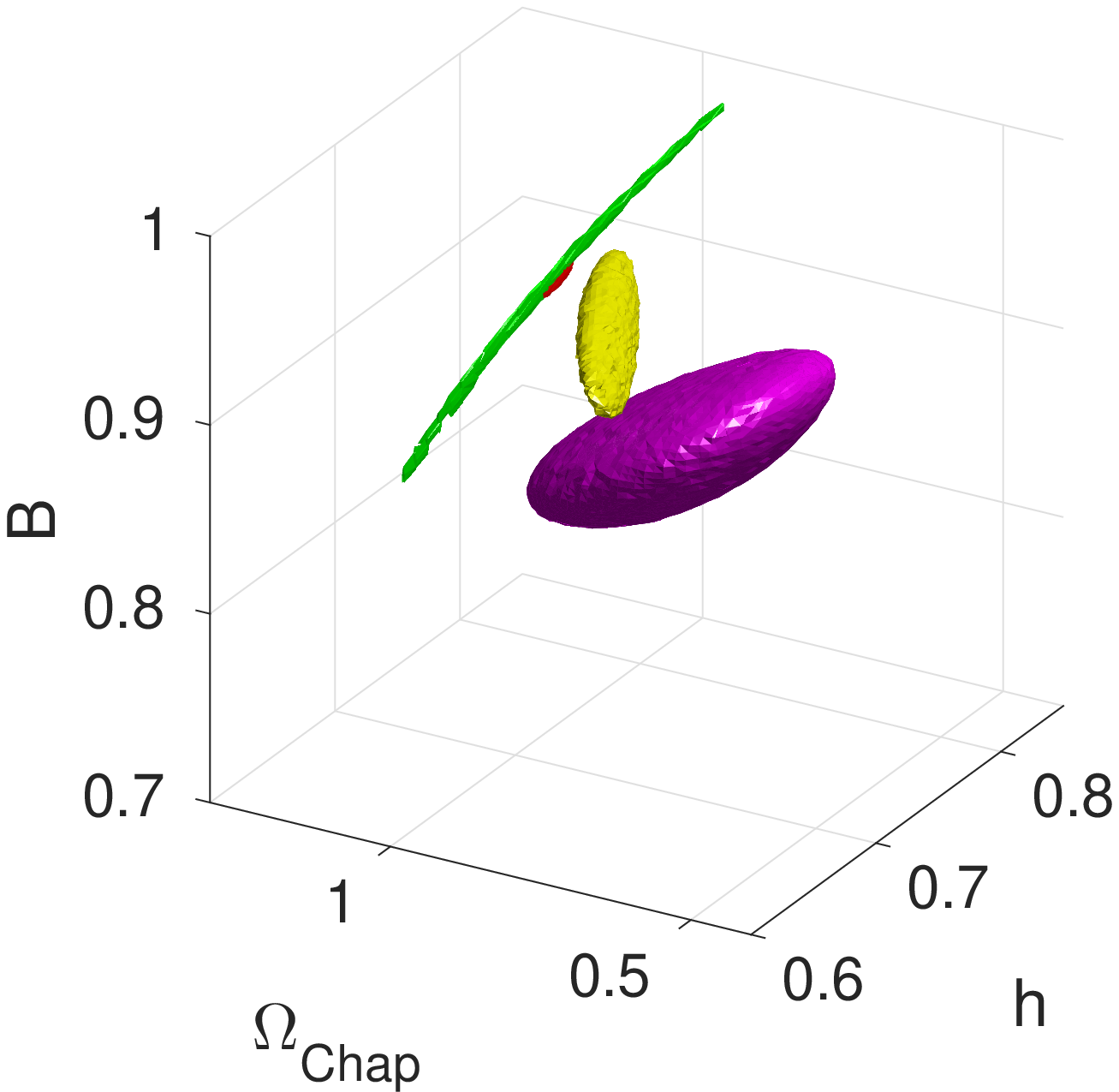}
\put(-490,170){(c) \hspace{20pt} $\alpha = 2/3$}
\put(-220,170){(d) \hspace{20pt} $\alpha = 1$}
\end{minipage}
\end{minipage}
\vspace*{-10pt}
\end{center}
\caption{\label{Fig:Chi2_3D}
The concordance between the three cosmological data sets is presented.
The 1$\sigma$ confidence domains are shown for the four Chaplygin gas models
with fixed parameter $\alpha=0$, $1/3$, $2/3$, and 1.
The axes denote the parameters that are varied in the MCMC sequences,
that is $h$, $\Omega_{\hbox{\scriptsize Chap}}$, and $B$.
The domain belonging to the supernovae data is shown in magenta,
the BAO domain in yellow, the CMB domain in green, and the combined $\chi^2$
search in red.
The projections are the same in all panels.
}
\end{figure}

As the subsection \ref{sec:Lya} has shown, the most stringent power
of the BAO data stems from the inclusion of the Lyman $\alpha$ forest
derived BAO data around the redshift $z\simeq 2.34$.
The Lyman $\alpha$ forest data consists of four data points
which belong to $d_A(z)/{r_s}$ and $d_H(z)/{r_s}$ obtained
from the cross- and the auto-correlation approach
each given in table \ref{Tab:BAO_data}.
Omitting these data around $z\simeq 2.34$ reduces
the power of the BAO data set.
To demonstrate that, the table \ref{Tab:Chi2_data_without_Lya} shows
the results obtained from MCMC sequences which use only the BAO data
without the data around $z\simeq 2.34$,
but the supernovae and CMB data are the same in the case of the joint analysis.
While $\chi^2_{\hbox{\scriptsize BAO}}$ takes on its minimum around
$\alpha\simeq -0.52$ (not listed in table \ref{Tab:Chi2_data})
by including the $z\simeq 2.34$ data,
that minimum is shifted to $\alpha\simeq +1.4$ by omitting the high $z$ data,
although this minimum belongs to an unrealistically large value for
the Hubble constant.
Furthermore, as the comparison of the column for $\chi^2_{\hbox{\scriptsize BAO}}$
of table \ref{Tab:Chi2_data} with that of table \ref{Tab:Chi2_data_without_Lya}
shows,
the dependence of $\chi^2_{\hbox{\scriptsize BAO}}$ on $\alpha$ is rather weak
without the $z\simeq 2.34$ data.
Table \ref{Tab:Chi2_data_without_Lya} reveals the modest increase by one
for $\chi^2_{\hbox{\scriptsize BAO}}$ by changing $\alpha=0$ to $\alpha=1$,
which has to be compared with the increase of 18.3 by using the
Lyman $\alpha$ data.
Let us now turn to the joint analysis,
where the MCMC sequences take the BAO, supernovae, and CMB data
simultaneously into account.
Here the comparison of the columns $\bar\chi^2_{\hbox{\scriptsize BAO}}$
of tables \ref{Tab:Chi2_data} and \ref{Tab:Chi2_data_without_Lya} shows
that the contribution $\bar\chi^2_{\hbox{\scriptsize BAO}}$ from
the BAO data to the minimum $\chi^2_{\hbox{\scriptsize BAO+SN+CMB}}$ is reduced
by more than ten without the $z\simeq 2.34$ data
although only four data points are left out.
This demonstrates the tension of the $z\simeq 2.34$ data with the
Chaplygin gas cosmology.
It was already found in \cite{Delubac_et_al_2015} that the auto-correlation
derived DR11 values of the Lyman $\alpha$ forest lead to values
that deviate by 7\% from the predictions of the $\Lambda$CDM model.
Thus the tension is also present in the $\Lambda$CDM model,
but the Chaplygin gas cosmology provides no remedy in this respect.

The value $\alpha=0$ is preferred by the MCMC sequences
which use the joint data sets $\chi^2_{\hbox{\scriptsize BAO+SN+CMB}}$ as seen in
tables \ref{Tab:Chi2_data} and \ref{Tab:Chi2_data_without_Lya}.
Let us now turn to table \ref{Tab:Chi2_data},
where the $z\simeq 2.34$ data are included.
Here, the least difference
$\bar\chi^2_{\hbox{\scriptsize BAO}}-\chi^2_{\hbox{\scriptsize BAO}}$
for the BAO data occurs for $\alpha=1/3$,
while the smallest difference
$\bar\chi^2_{\hbox{\scriptsize SN}}-\chi^2_{\hbox{\scriptsize SN}}$ due to the
supernovae data happens at $\alpha=0$.
The CMB data show only a modest increase by one for $\alpha=1/3$ and
$\alpha=2/3$.
So one concludes that although the total $\chi^2_{\hbox{\scriptsize BAO+SN+CMB}}$
value is minimal for $\alpha=0$, 
the tension between the BAO and CMB data sets is reduced for
$\alpha$ around 1/3.

In order to visualise these dependencies,
the figure \ref{Fig:Chi2_3D} shows the 1$\sigma$ confidence domains
generated by the various MCMC sequences for the four fixed values
$\alpha=0$, $1/3$, $2/3$, and 1.
Here and in the following,
the BAO data are used including the $z\simeq 2.34$ data.
The 1$\sigma$ confidence domains are chosen instead of, e.\,g.\
2$\sigma$ confidence domains, because of their reduced volume and, thus,
they show more clearly the shifting in the parameter space.
The domains belonging to the CMB chains (shown in green) extend to
large values of the Hubble constant $h$ and, for clarity, are truncated
at $h=0.85$ in figure \ref{Fig:Chi2_3D}.
The domains belonging to the supernovae chains are shown in magenta and
those of the BAO chains in yellow, while the confidence domains of the
total MCMC chains are plotted in red.

\begin{figure}
\begin{center}
\hspace*{-50pt}\begin{minipage}{20cm}
\vspace*{-30pt}
\begin{minipage}{20cm}
\hspace*{-40pt}\includegraphics[width=10.0cm]{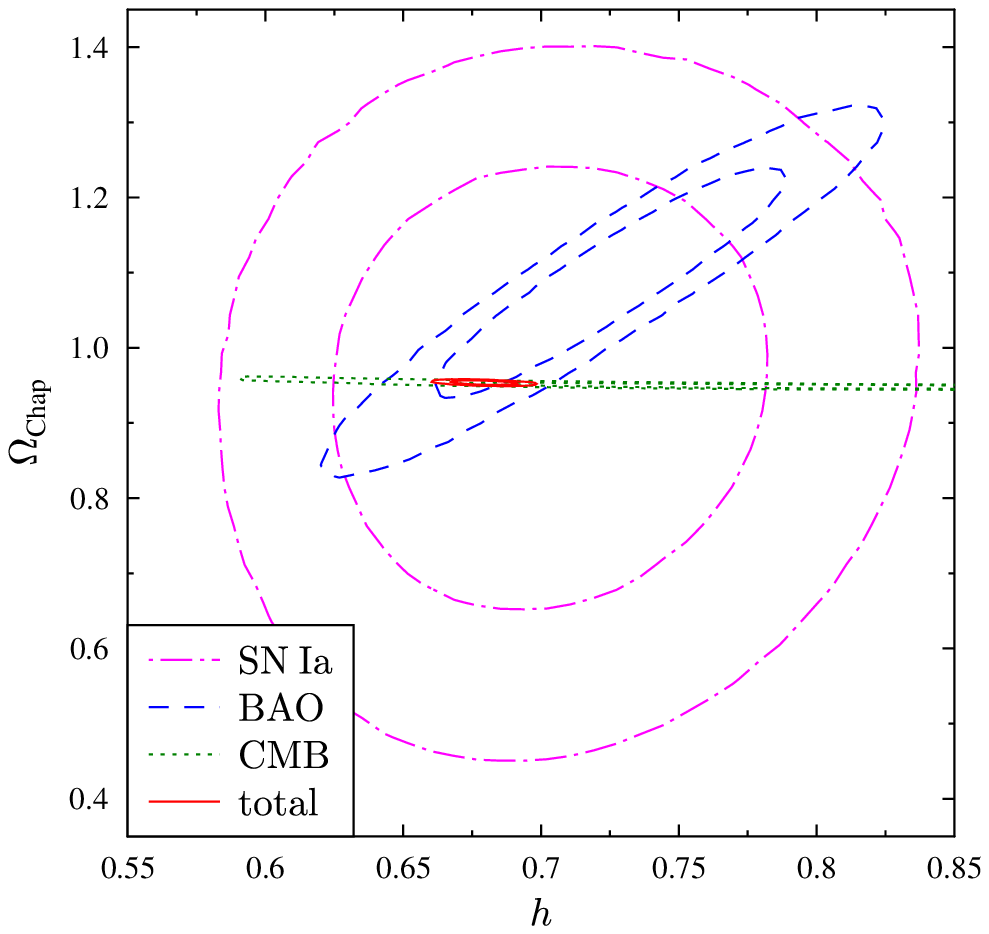}
\hspace*{-40pt}\includegraphics[width=10.0cm]{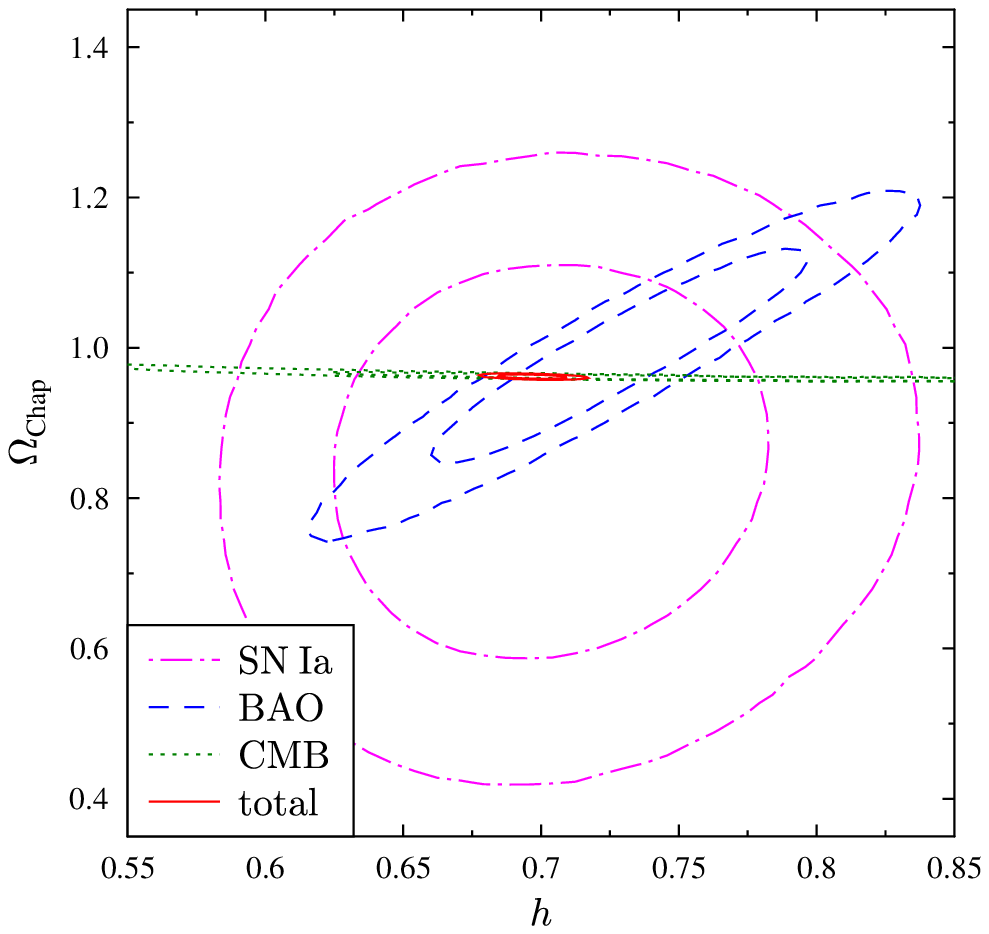}
\put(-475,220){(a) \hspace{40pt} $\alpha = 0$}
\put(-220,220){(b) \hspace{20pt} $\alpha = 1/3$}
\vspace*{-50pt}
\end{minipage}
\begin{minipage}{20cm}
\hspace*{-40pt}\includegraphics[width=10.0cm]{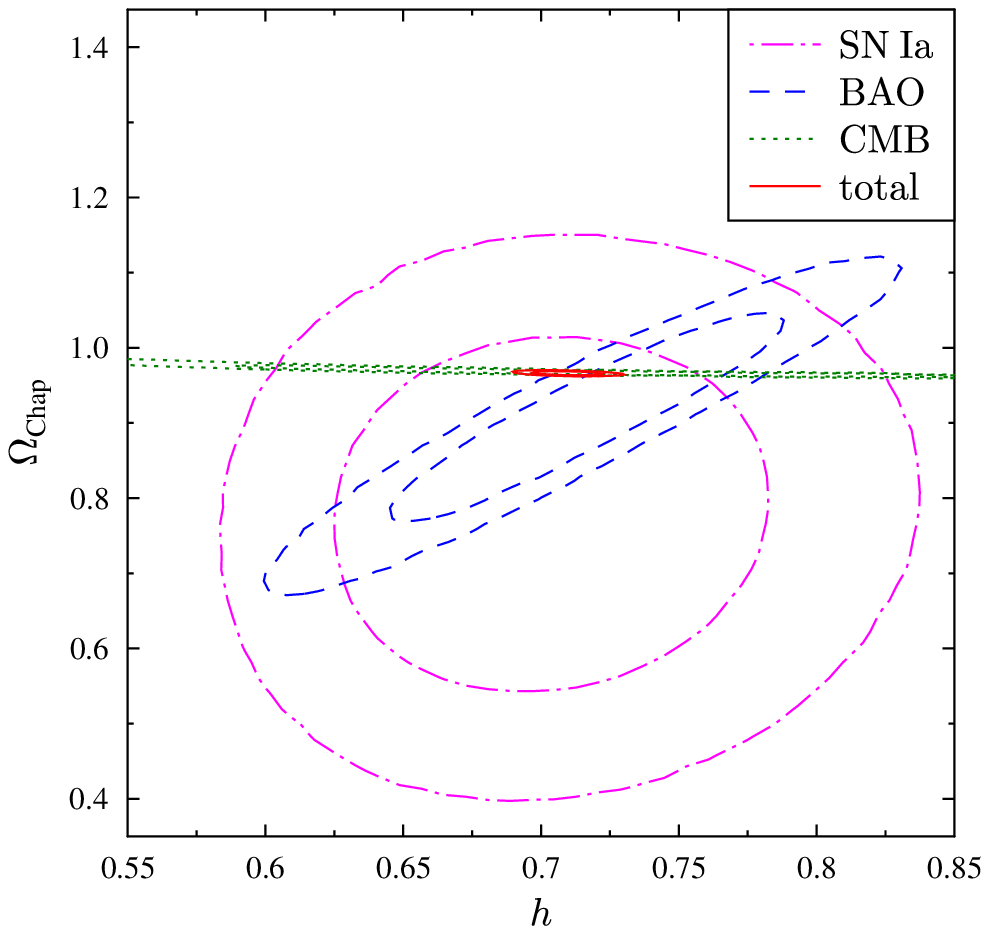}
\hspace*{-40pt}\includegraphics[width=10.0cm]{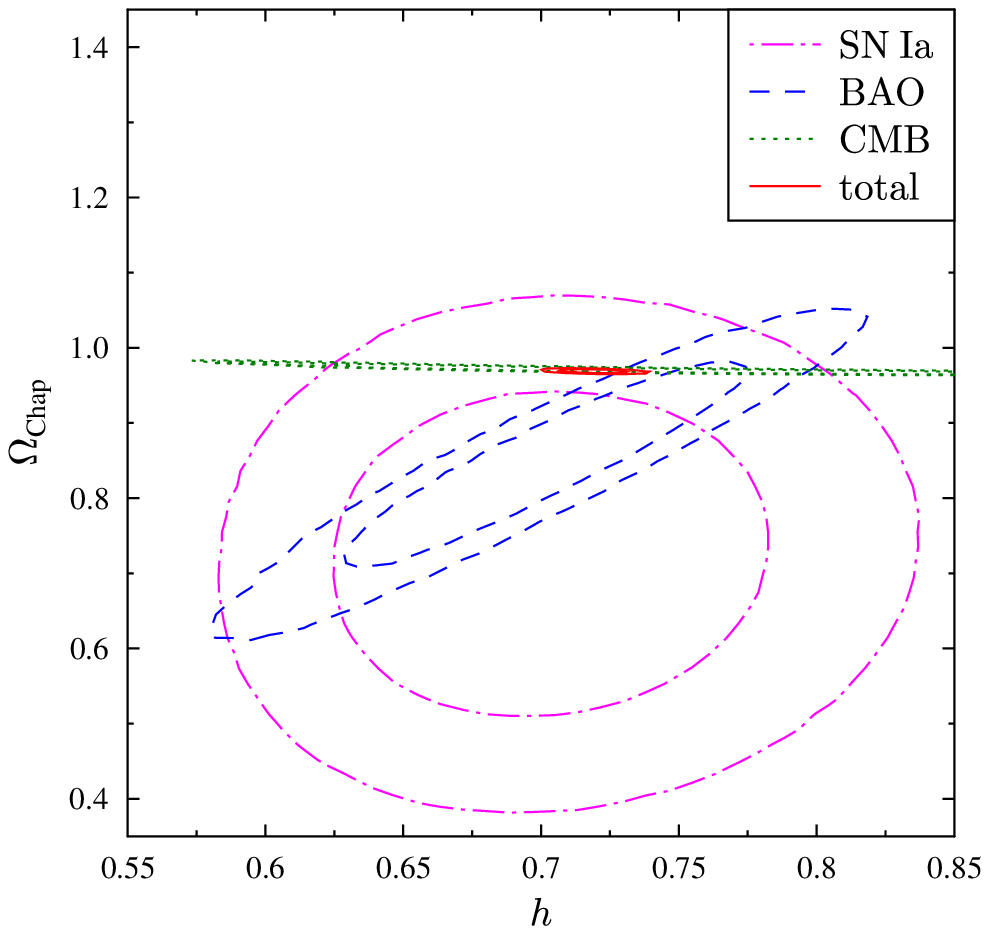}
\put(-475,220){(c) \hspace{20pt} $\alpha = 2/3$}
\put(-220,220){(d) \hspace{20pt} $\alpha = 1$}
\end{minipage}
\end{minipage}
\vspace*{-30pt}
\end{center}
\caption{\label{Fig:Chi2_2D_h_OChap}
The 1$\sigma$ and 2$\sigma$ confidence contours are shown for the four
Chaplygin gas models with fixed parameter $\alpha=0$, $1/3$, $2/3$, and 1.
The confidence domains are projected onto the
$h-\Omega_{\hbox{\scriptsize Chap}}$ plane.
To emphasise the shift of the confidence domains depending on $\alpha$,
all four panels use the same scaling at the axes.
}
\end{figure}

\begin{figure}
\begin{center}
\hspace*{-50pt}\begin{minipage}{20cm}
\vspace*{-30pt}
\begin{minipage}{20cm}
\hspace*{-40pt}\includegraphics[width=10.0cm]{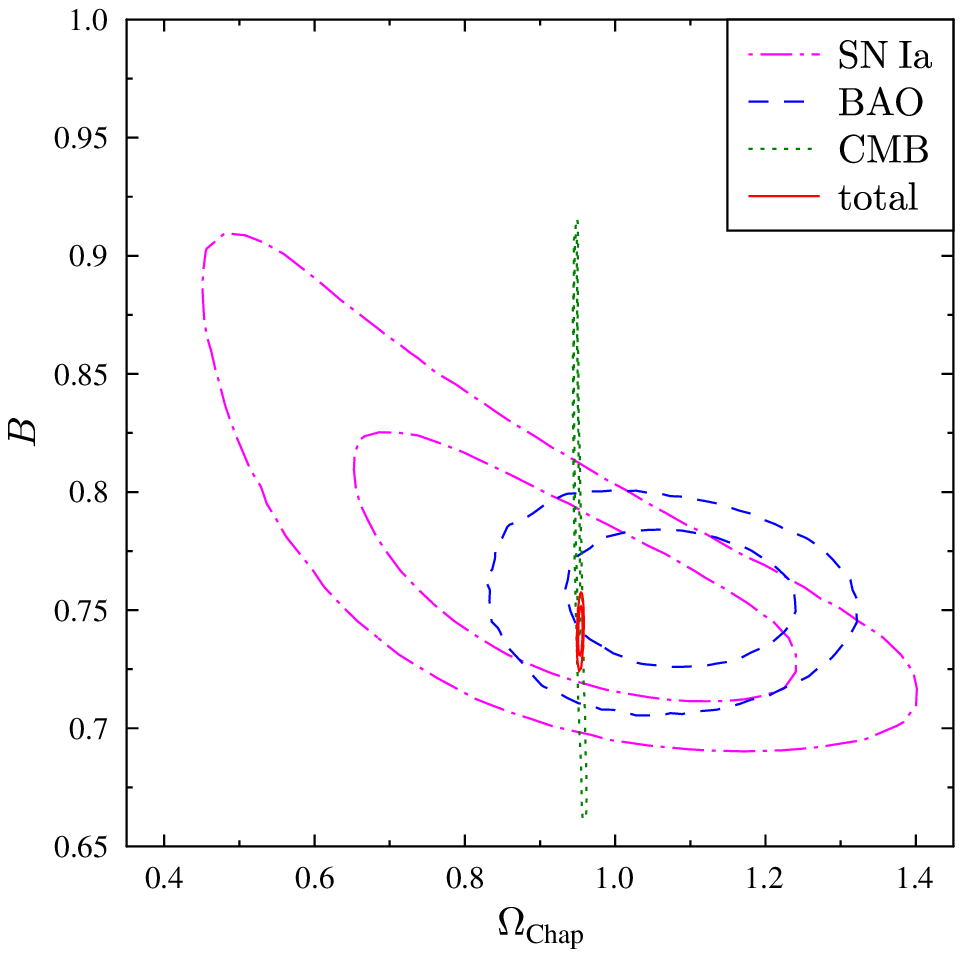}
\hspace*{-40pt}\includegraphics[width=10.0cm]{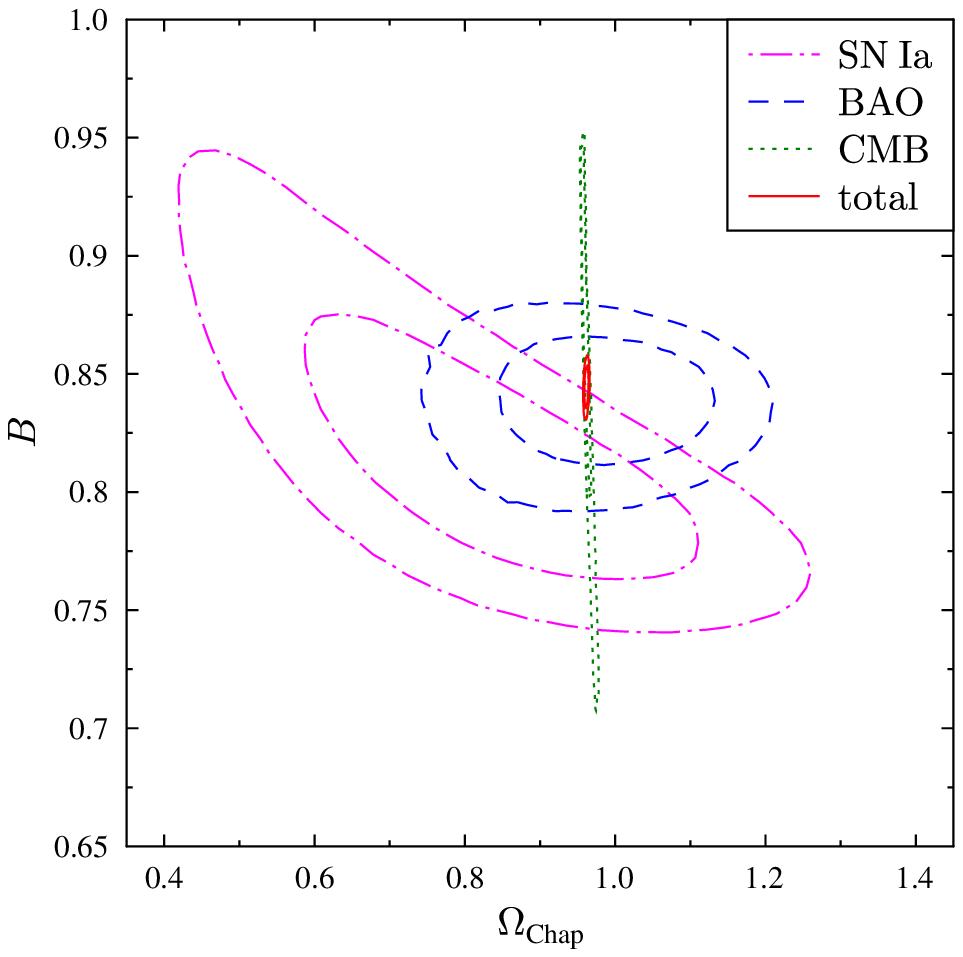}
\put(-475,220){(a) \hspace{20pt} $\alpha = 0$}
\put(-220,220){(b) \hspace{20pt} $\alpha = 1/3$}
\vspace*{-50pt}
\end{minipage}
\begin{minipage}{20cm}
\hspace*{-40pt}\includegraphics[width=10.0cm]{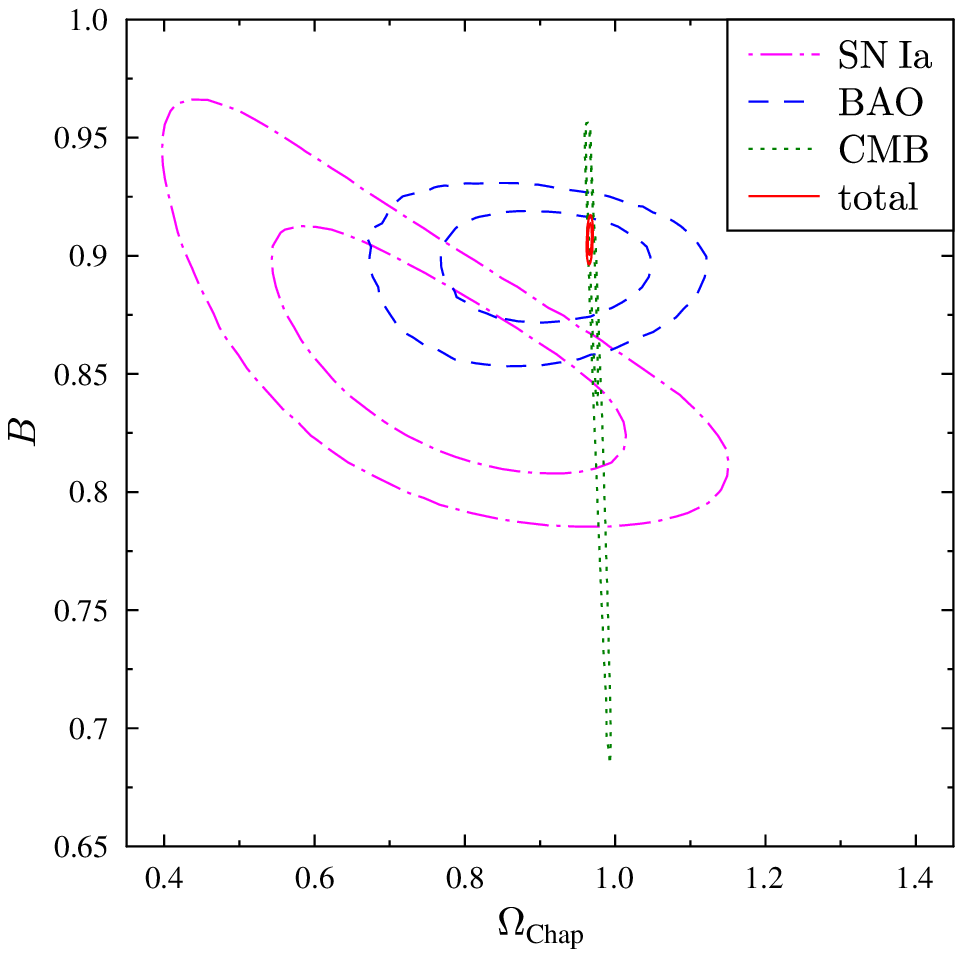}
\hspace*{-40pt}\includegraphics[width=10.0cm]{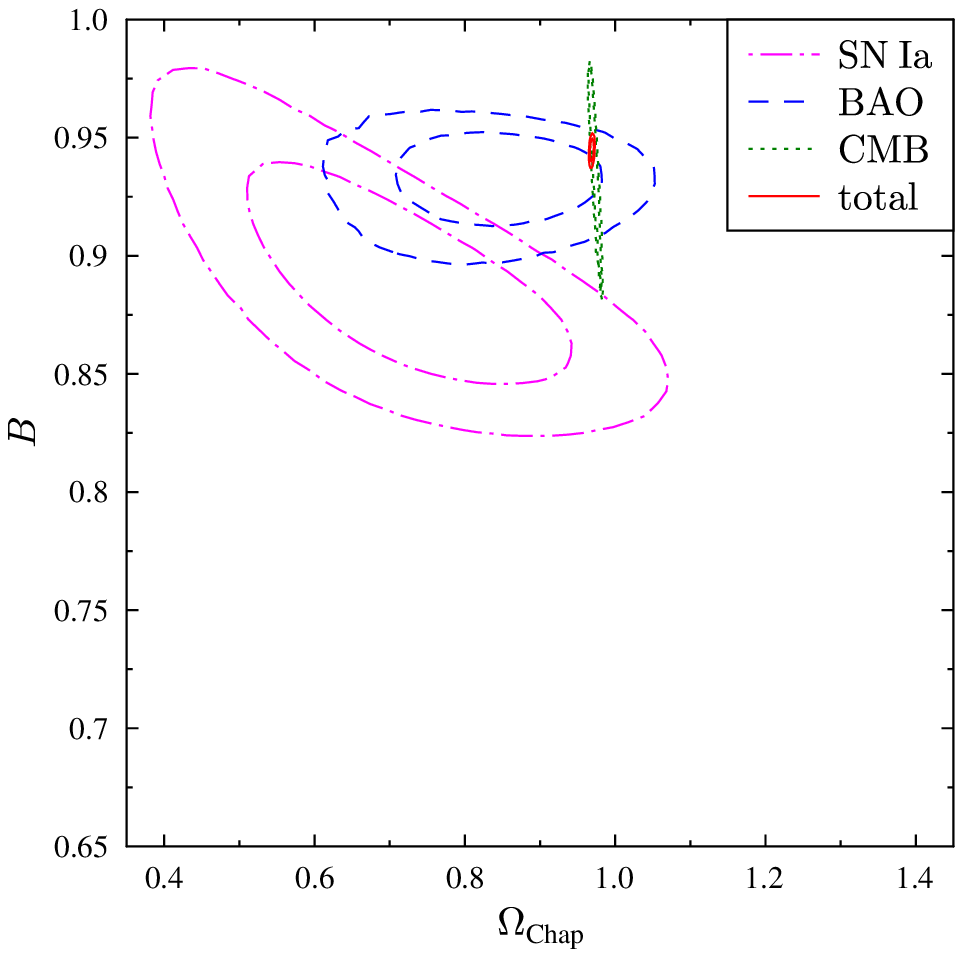}
\put(-475,80){(c) \hspace{20pt} $\alpha = 2/3$}
\put(-220,80){(d) \hspace{20pt} $\alpha = 1$}
\end{minipage}
\end{minipage}
\vspace*{-30pt}
\end{center}
\caption{\label{Fig:Chi2_2D_OChap_B}
The 1$\sigma$ and 2$\sigma$ confidence contours are shown for the four
Chaplygin gas models with fixed parameter $\alpha=0$, $1/3$, $2/3$, and 1.
The confidence domains are projected onto the
$\Omega_{\hbox{\scriptsize Chap}}-B$ plane.
As in figure \ref{Fig:Chi2_2D_h_OChap},
all four panels use the same scaling at the axes.
}
\end{figure}

\begin{figure}
\begin{center}
\begin{minipage}{10cm}
\vspace*{-30pt}\includegraphics[width=10.0cm]{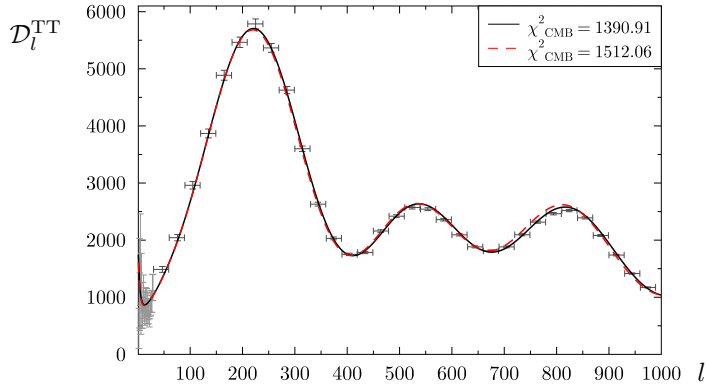}
\put(-290,147){${\cal D}_l^{\hbox{\scriptsize TT}}$}
\put(-30,19){$l$}
\end{minipage}
\vspace*{-10pt}
\end{center}
\caption{\label{Fig:cmb_comparison_alpha_0_chi2}
Two angular power spectra ${\cal D}_l^{\hbox{\scriptsize TT}}$ are shown
for $\alpha=0$,
which possess different values of $\chi^2_{\hbox{\scriptsize CMB}}$.
The full curve belongs to the model with the lowest value of
$\chi^2_{\hbox{\scriptsize CMB}}$ for the case $\alpha=0$
($h=0.8859$, $\Omega_{\hbox{\scriptsize bar}}=0.0284$,
$\Omega_{\hbox{\scriptsize Chap}}=0.9476$, $B=0.8471$)
while the dashed curve belongs to a model with a $\chi^2_{\hbox{\scriptsize CMB}}$
value which is higher by 121
($h=0.8570$, $\Omega_{\hbox{\scriptsize bar}}=0.0304$,
$\Omega_{\hbox{\scriptsize Chap}}=0.9488$, $B=0.8309$).
The binned data of the Planck 2015 data release
\cite{Planck_2015_I} are plotted for $l>30$
with the corresponding error bars,
while up to $l=29$ all data points are shown.
}
\end{figure}

For $\alpha=0$ the red confidence domain of the total
$\chi^2_{\hbox{\scriptsize BAO+SN+CMB}}$ MCMC sequences is within that of
the supernovae close to the common intersection point
of the three 1$\sigma$ confidence domains each belonging to a single data set.
This concordance between the data sets is removed with increasing $\alpha$.
For $\alpha=1/3$ there is no common overlap between the three individual
1$\sigma$ confidence domains, although each pair of them possesses one.
This gets worse for $\alpha=2/3$ as seen in figure \ref{Fig:Chi2_3D}(c),
where the CMB and the supernovae domains do not overlap at all and
those of the supernovae and BAO are separated by a tiny gap.
For $\alpha=1$ the separation between the confidence domains increases further.
It is worthwhile to note
that the red confidence domain tends to be close to that of the CMB domain,
which is a consequence of the constraining power of the Planck Likelihood Code.
The inspection of the ${\cal D}_l^{\hbox{\scriptsize TT}}$ spectra for the
Chaplygin gas cosmology reveals
that a very steep increase of the $\chi^2_{\hbox{\scriptsize CMB}}$ values
occurs due to a slight mismatch of the second and third acoustic peak
compared to those measured by Planck.
In order to illustrate this sensitivity,
figure \ref{Fig:cmb_comparison_alpha_0_chi2} displays the
angular power spectra ${\cal D}_l^{\hbox{\scriptsize TT}}$ of two models
whose $\chi^2_{\hbox{\scriptsize CMB}}$ values differ by 121.
A difference in ${\cal D}_l^{\hbox{\scriptsize TT}}$ is only appreciable
at the third acoustic peak.

In figures \ref{Fig:Chi2_2D_h_OChap} and \ref{Fig:Chi2_2D_OChap_B},
the projections of the 1$\sigma$ confidence domains of the three-dimensional
plots of figure \ref{Fig:Chi2_3D} are shown on a two parameter plane and,
in addition, the 2$\sigma$ confidence domains are also plotted.
This perspective allows a more quantitative comparison of the
confidence domains, especially how far they overlap.
The figure \ref{Fig:Chi2_2D_h_OChap} displays these domains in the 
$h-\Omega_{\hbox{\scriptsize Chap}}$ plane.
As already noted in connection with figure \ref{Fig:Chi2_3D},
the confidence domains of the CMB data are very thin bands
which are projected onto the horizontal bands in
figure \ref{Fig:Chi2_2D_h_OChap}.
Comparing the  $1\sigma$ confidence domains belonging to
the joint analysis of the four panels in figure \ref{Fig:Chi2_2D_h_OChap},
one observes that the Hubble constant $h$ increases with $\alpha$.
This is a consequence of the BAO data as a comparison of the
four panels in figure \ref{Fig:Chi2_2D_h_OChap} reveals.
The confidence domains belonging to the BAO data move to lower values
of $\Omega_{\hbox{\scriptsize Chap}}$ with increasing $\alpha$
and due to their inclination within the $h-\Omega_{\hbox{\scriptsize Chap}}$ plane,
the intersection with the CMB contours occurs at higher values of $h$.
Figure \ref{Fig:Chi2_2D_OChap_B},
where the $\Omega_{\hbox{\scriptsize Chap}}-B$ plane is plotted,
shows that $B$ also increases with $\alpha$.
This behaviour stems again from the BAO data, since the CMB data lead
to the vertical bands in figure \ref{Fig:Chi2_2D_OChap_B} allowing
a wide range of values of $B$.
Because of $w_{\hbox{\scriptsize Chap}}(z=0)=-B$,
the current equation of state (\ref{Eq:Chaplygin_equation_of_state})
tends towards more negative values.
If the values of $w_{\hbox{\scriptsize Chap}}(z=0)$ are to be compared with the
$\Lambda$CDM model, one has to combine the equation of state of
the CDM component with that of the cosmological constant leading to
$w_{\Lambda\hbox{\scriptsize CDM}}(z=0)=
-1/(1+\Omega_{\hbox{\scriptsize CDM}}/\Omega_\Lambda)$.

\begin{figure}
\begin{center}
\begin{minipage}{10cm}
\vspace*{-30pt}\includegraphics[width=10.0cm]{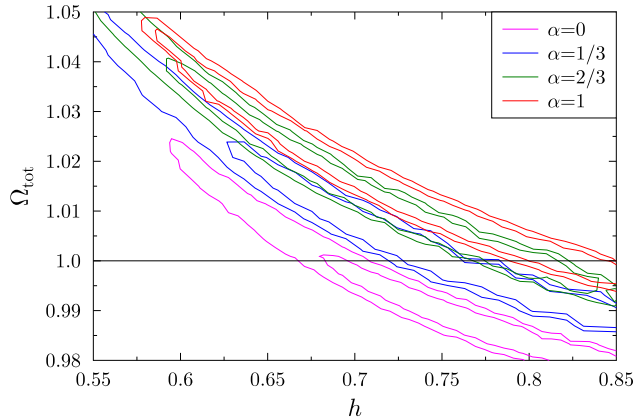}
\end{minipage}
\vspace*{-10pt}
\end{center}
\caption{\label{Fig:Chi2_2D_h_Otot}
The 1$\sigma$ and 2$\sigma$ confidence contours computed from the CMB data
are shown projected onto the $h-\Omega_{\hbox{\scriptsize tot}}$ plane.
The confidence contours belong to the four Chaplygin gas models
with a fixed parameter $\alpha=0$, $1/3$, $2/3$, and 1.
The horizontal line indicates spatially flat models with
$\Omega_{\hbox{\scriptsize tot}}=1$.
}
\end{figure}

The CMB data point to more or less spatially flat models since
$\Omega_{\hbox{\scriptsize tot}}$, equation (\ref{Def:Omega_tot}),
turns out to be of order one.
Figure \ref{Fig:Chi2_2D_h_Otot} shows that the confidence domains
projected onto the $h-\Omega_{\hbox{\scriptsize tot}}$ plane
shift to larger values of $\Omega_{\hbox{\scriptsize tot}}$ with increasing $\alpha$,
which is the reverse behaviour as it is observed in the case of the BAO data.
But the shift caused by the CMB data is so small
that the confidence domains overlap for neighbouring values of $\alpha$.
Note also, that only the small interval $[0.98,1.05]$ for 
$\Omega_{\hbox{\scriptsize tot}}$ is displayed.
Furthermore, the figure demonstrates that a possible restriction
to spatially flat models in the MCMC sequences leads
to larger values of the Hubble constant $h$ with increasing $\alpha$.
Without such a restriction, the confidence domains for $\alpha=0$
point to a slightly negative curvature,
while for $\alpha=2/3$ and  $\alpha=1$ a positive curvature is preferred,
unless unrealistically large values of $h$ are accepted.

\subsection{MCMC analysis with varying values of $\alpha$}
\label{sec:MCMC_with_varying_alpha}

In the previous subsection \ref{sec:MCMC_with_fixed_alpha},
the results are presented for MCMC sequences
in which the equation of state parameter $\alpha$ is held fixed.
This choice is motivated by the fact that $\alpha=1$ corresponds to the
originally proposed Chaplygin gas cosmology,
while $\alpha=0$ leads to the background cosmology of the
$\Lambda$CDM model.
So these two $\alpha$ values and two intermediate values are chosen for
the analysis which shows that $\alpha=0$ is superior compared to the other
three values of $\alpha$
if all three data sets are simultaneously taken into account.
Since the differences in these $\alpha$ values are relatively large,
a further MCMC sequence is generated with 150\,000 iterations,
which varies also the parameter $\alpha$ in addition to the previously
varied parameters $h$, $\Omega_{\hbox{\scriptsize Chap}}$, and $B$.
The parameter $\alpha$ is restricted to $\alpha>-1$ in the MCMC sequence.
In that way, the distribution of the $\alpha$ values can be inferred.
In this section, the analysis takes all BAO data given in
table \ref{Tab:BAO_data} as well as the above supernovae and CMB data
into account.
The best-fit Chaplygin model obtained in this MCMC sequence has
$\chi^2_{\hbox{\scriptsize BAO+SN+CMB}}=1977.53$,
which is only marginally smaller than the value in table \ref{Tab:Chi2_data}
for $\alpha=0$.

\begin{figure}
\begin{center}
\hspace*{-50pt}\begin{minipage}{20cm}
\vspace*{-20pt}
\begin{minipage}{20cm}
\hspace*{-50pt}\includegraphics[width=10.0cm]{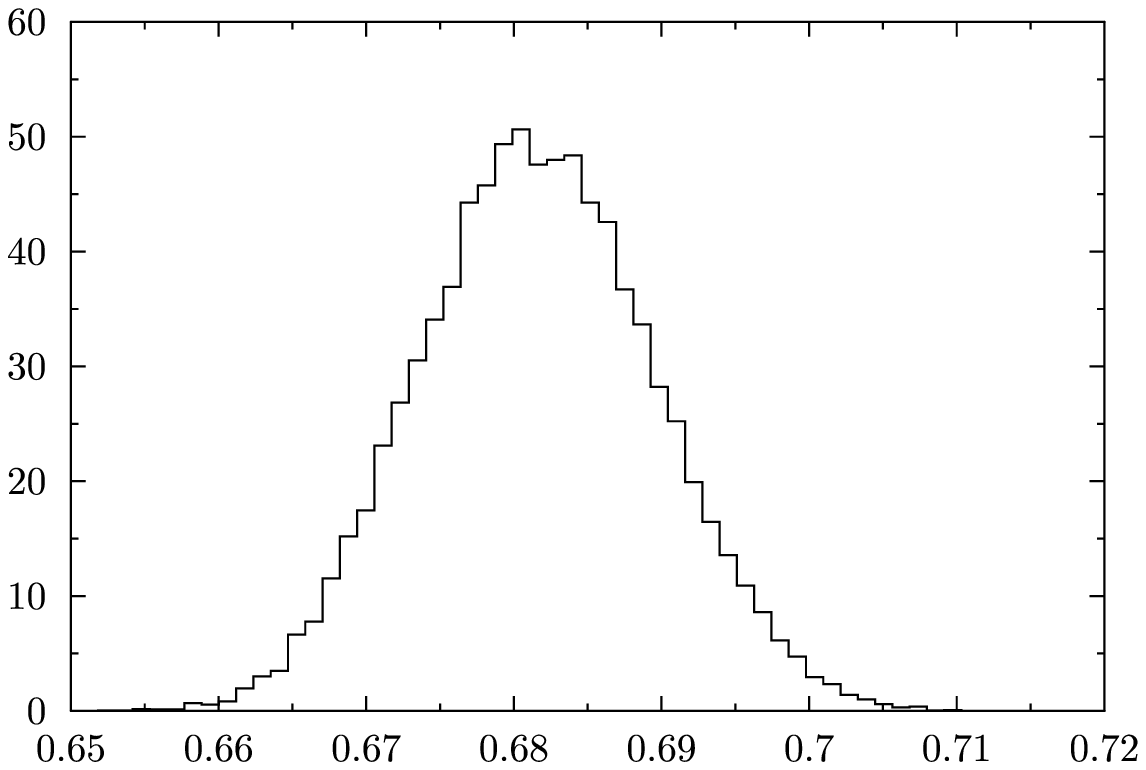}
\put(-225,135){(a) \hspace{6pt} $h$}
\put(-33,19){$h$}
\hspace*{-40pt}\includegraphics[width=10.0cm]{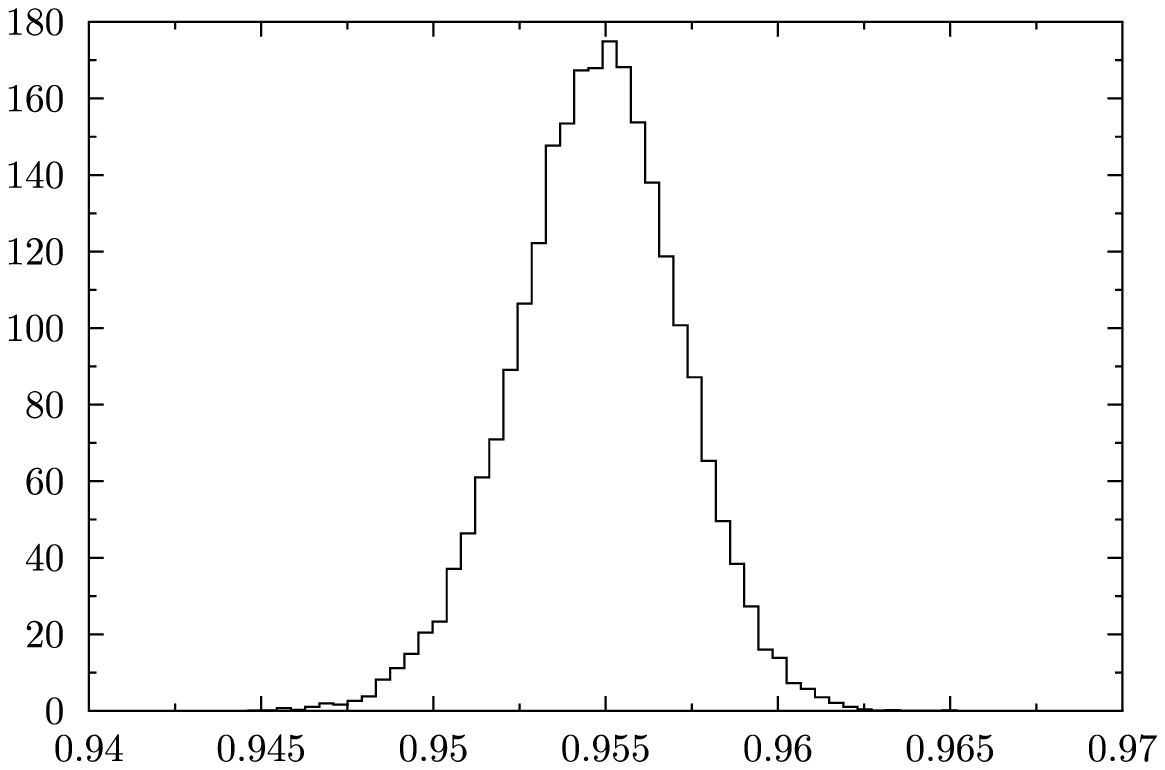}
\put(-225,135){(b) \hspace{6pt} $\Omega_{\hbox{\scriptsize Chap}}$}
\put(-35,19){$\Omega_{\hbox{\scriptsize Chap}}$}
\vspace*{-20pt}
\end{minipage}
\begin{minipage}{20cm}
\hspace*{-50pt}\includegraphics[width=10.0cm]{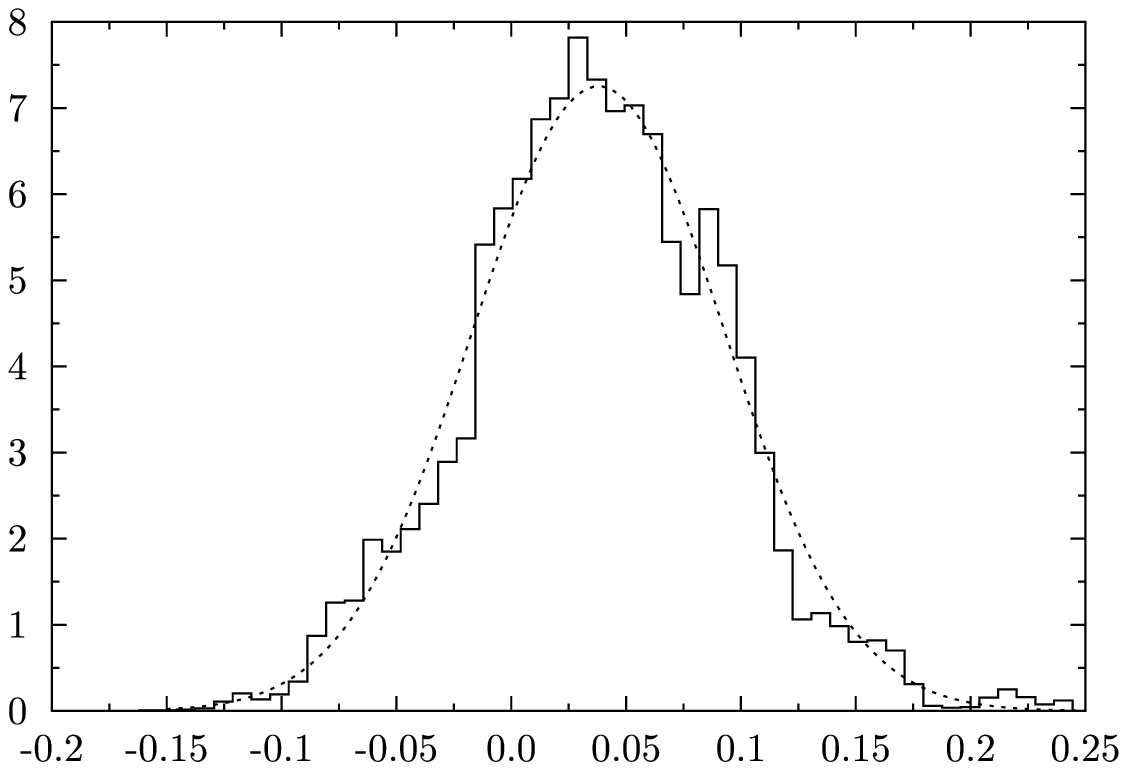}
\put(-225,135){(c) \hspace{6pt} $\alpha$}
\put(-33,19){$\alpha$}
\hspace*{-40pt}\includegraphics[width=10.0cm]{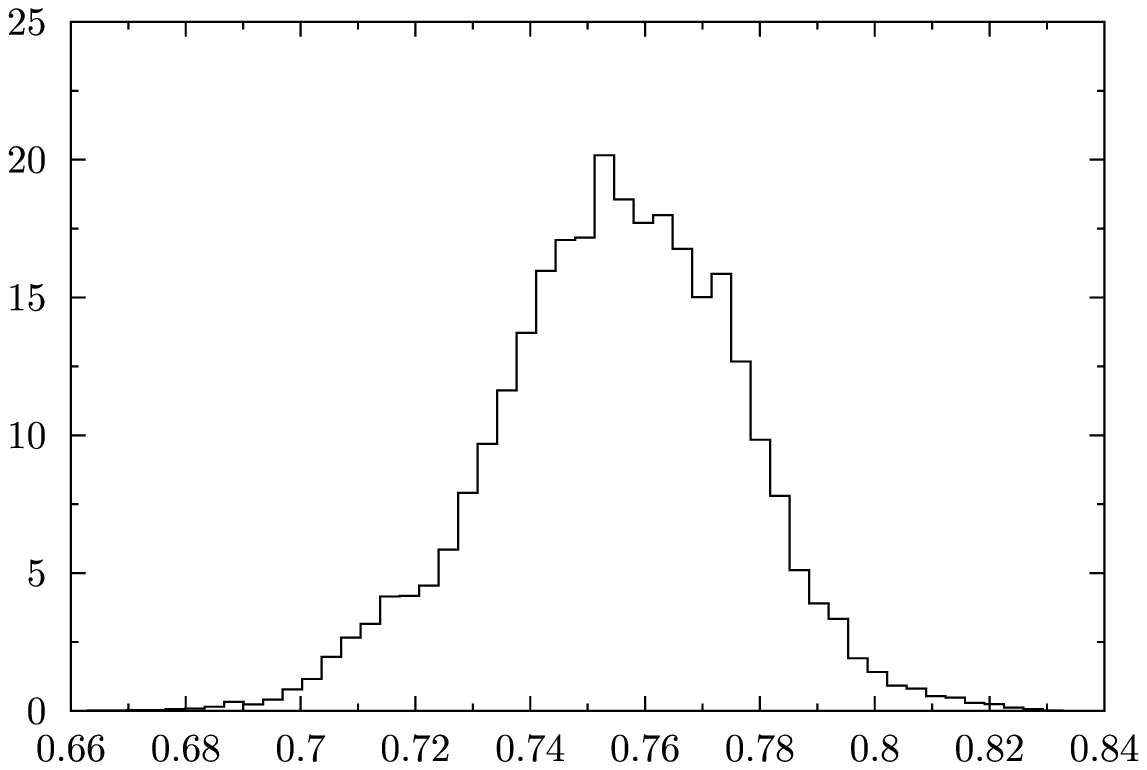}
\put(-225,135){(d) \hspace{6pt} $B$}
\put(-35,19){$B$}
\end{minipage}
\end{minipage}
\vspace*{-20pt}
\end{center}
\caption{\label{Fig:hist_alpha_distribution}
The distribution of the parameters
derived from the MCMC sequence with varying values of $\alpha$.
In panel (c) the distribution of the $\alpha$ parameter is compared to
a normal distribution with mean $\bar\alpha = 0.038$ and width $0.055$.
The value $\alpha = 0$ is thus within the 1$\sigma$ confidence interval,
but the best-fit models are close to $\alpha = 0.04$.
}
\end{figure}

The obtained distributions for the four parameters are shown in
figure \ref{Fig:hist_alpha_distribution}.
The mean values and their $1\sigma$ deviations are
$h=0.6816\pm 0.0079$,
$\Omega_{\hbox{\scriptsize Chap}}=0.9547\pm 0.0024$,
$\alpha=0.038\pm 0.055$, and
$B=0.755\pm 0.021$.
The obtained value of the Hubble constant $h$ is in agreement with the
low value found by Planck \cite{Planck_2015_I}.
It is worthwhile to note that this does not arise from the supernovae data
since the last term in (\ref{Def:Chi2_Sn}) is omitted in this joint analysis,
and thus, there is no prior in $h$.
As stated below (\ref{Def:Chi2_Sn}) this term is only taken into account
when the supernovae data are used alone.
The $\Lambda$CDM model, which corresponds to $\alpha=0$, is within the
$1\sigma$ confidence interval
although the peak of the distribution is shifted to larger values of $\alpha$.
The value of $\Omega_{\hbox{\scriptsize Chap}}$ is also in the range
expected from the $\Lambda$CDM model.
Assuming $\Omega_{\hbox{\scriptsize CDM}}\simeq 0.25$ and $\Omega_\Lambda\simeq 0.7$,
the dark component contribution of the $\Lambda$CDM model is of the
order 0.95 which agrees with $\Omega_{\hbox{\scriptsize Chap}}$.
The dark components of the $\Lambda$CDM model lead to the effective
equation of state
$w_{\Lambda\hbox{\scriptsize CDM}}(z=0)= -1/(1+\Omega_{\hbox{\scriptsize CDM}}/\Omega_\Lambda)$
which is thus of the order $w_{\Lambda\hbox{\scriptsize CDM}}(z=0)\simeq -0.74$
which in turn lies within the corresponding $1\sigma$ confidence interval
for $w_{\hbox{\scriptsize Chap}}(z=0)=-B$.
The total density is obtained as $\Omega_{\hbox{\scriptsize tot}}=1.0028\pm 0.0026$
which thus shows a tendency towards positive curvature.

\begin{figure}
\begin{center}
\begin{minipage}{10cm}
\vspace*{-30pt}\includegraphics[width=10.0cm]{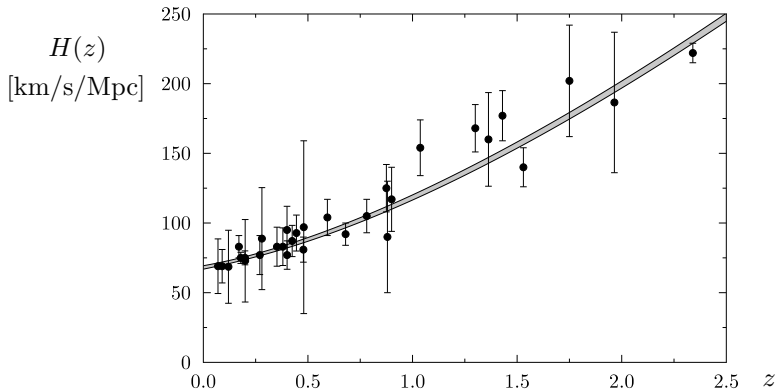}
\put(-300,145){$H(z)$}
\put(-316,130){[km/s/Mpc]}
\put(-30,20){$z$}
\end{minipage}
\vspace*{-10pt}
\end{center}
\caption{\label{Fig:Hubble_parameter_variation}
The variation of the Hubble parameter $H(z)$ is shown as a grey band
for those Chaplygin gas cosmologies
which are within the $1\sigma$ confidence domain
with respect to $\chi^2_{\hbox{\scriptsize BAO+SN+CMB}}$.
The observational Hubble data are taken from the compilation
\cite{Moresco_et_al_2016}.
The Lyman $\alpha$ forest derived Hubble parameter at $z=2.34$ is also
plotted.
}
\end{figure}

Figure \ref{Fig:Hubble_parameter_variation} shows the variation of
the Hubble parameter $H(z)$ for redshifts below $z=2.5$.
The distribution shown as a grey band is computed from
all Chaplygin gas cosmologies
which are within the $1\sigma$ confidence domain according to the
joint analysis.
Again, the MCMC models of the sequence with a varying parameter $\alpha$
are used.
The figure also shows the observational Hubble data,
which are taken from the compilation of \cite{Moresco_et_al_2016}.
In addition, the Hubble parameter at $z=2.34$ derived from the
Lyman $\alpha$ data is plotted \cite{Delubac_et_al_2015}.
The Chaplygin gas values match very well to the observations
and their variation is much smaller than the $1\sigma$ errors
of the data.

\section{Summary}
\label{sec:Summary}

This paper discusses whether a better description of recent cosmological
observations can be achieved,
when the dark sector of the $\Lambda$CDM concordance model is replaced
by the generalised Chaplygin gas.
This prototypical model for a unified dark matter model with the
equation of state $p = -A/\varepsilon^\alpha$ has the advantage
that the $\Lambda$CDM model is recovered for $\alpha=0$
at the background level.
It turns out that the CMB statistics is the same for the generalised Chaplygin
gas with $\alpha=0$ and the $\Lambda$CDM cosmology in linear
perturbation theory.
This allows to test whether there are models in the neighbourhood of
the concordance model that provide a better match to the data.

In this paper, the Planck 2015 cosmic microwave data, 
the SN Ia Union 2.1 compilation of the Supernova Cosmology Project,
and the BAO data of the Data Release 12 of the
SDSS-III/BOSS spectroscopic galaxy survey
are used for a joint analysis with respect to the Chaplygin gas cosmology.
Special emphasis is put on the four models with $\alpha=0$, $1/3$, $2/3$,
and 1.
As visualised in figure \ref{Fig:Chi2_3D},
a concordance between the three sets of cosmological data is only obtained for
values of $\alpha$ close to zero, where all three $1\sigma$ confidence
domains overlap.
Already for $\alpha=1/3$ this concordance is lost and,
at least at the $1\sigma$ level,
there is no common overlap between the three confidence domains.
The contribution of the BAO data to this behaviour depends
on the inclusion of the Lyman $\alpha$ forest derived BAO data,
as the comparison of table \ref{Tab:Chi2_data} with
table \ref{Tab:Chi2_data_without_Lya} reveals.
The strong increase of the $\chi^2$ values of the BAO data is observed
only by taking the Lyman $\alpha$ data into account.
However, a more modest increase of $\chi^2$ with increasing values of $\alpha$
is also observed for the CMB data,
while the supernovae data show no dependence on $\alpha$.

So the final conclusion is that the recent cosmological data favour values of
$\alpha$ very close to zero.
A joint analysis with varying values of $\alpha$ leads to
$\alpha=0.038\pm 0.055$,
which includes the $\Lambda$CDM model within the $1\sigma$ interval.

\section*{Acknowledgements}

The Planck 2015 data \cite{Planck_2015_I} obtained from the LAMBDA website
(\texttt{http:/$\!$/ lambda.gsfc.nasa.gov}),
the SN Ia Union 2.1 compilation of the Supernova Cosmology Project
\cite{Suzuki_et_al_2012},
and the SDSS-III/BOSS spectroscopic galaxy survey DR12 
\cite{Chuang_et_al_2016}
are used in this work.
The Planck Likelihood Code R2.00 provided on the Planck website
\texttt{http:/$\!$/pla.esac.esa.int/pla/\#cosmology}
is also used.


\section*{References}

\bibliography{../../bib_astro}

\begin{thebibliography}{10}
\expandafter\ifx\csname url\endcsname\relax
  \def\url#1{\texttt{#1}}\fi
\expandafter\ifx\csname urlprefix\endcsname\relax\def\urlprefix{URL }\fi
\expandafter\ifx\csname href\endcsname\relax
  \def\href#1#2{#2} \def\path#1{#1}\fi

\bibitem{Huterer_Shafer_2017}
D.~{Huterer}, D.~L. {Shafer}, {Dark energy two decades after: Observables,
  probes, consistency tests}, ArXiv e-prints\href
  {http://arxiv.org/abs/1709.01091} {\path{arXiv:1709.01091}}.

\bibitem{Buchert_Coley_Kleinert_Roukema_Wiltshire_2016}
T.~{Buchert}, A.~A. {Coley}, H.~{Kleinert}, B.~F. {Roukema}, D.~L. {Wiltshire},
  {Observational challenges for the standard FLRW model}, Int.\ J.\ Mod.\
  Phys.\ D 25 (2016) 1630007.
\newblock \href {http://arxiv.org/abs/1512.03313} {\path{arXiv:1512.03313}}.

\bibitem{Kamenshchik_Moschella_Pasquier_2001}
A.~{Kamenshchik}, U.~{Moschella}, V.~{Pasquier}, {An alternative to
  quintessence}, Physics Letters B 511 (2001) 265--268.
\newblock \href {http://arxiv.org/abs/gr-qc/0103004}
  {\path{arXiv:gr-qc/0103004}}.

\bibitem{Fabris_Goncalves_deSouza__2001}
J.~C. {Fabris}, S.~V.~B. {Goncalves}, P.~E. {de Souza}, {Density perturbations
  in an Universe dominated by the Chaplygin gas}, Gen.\ Rel.\ Grav. 34 (2002)
  53--63.
\newblock \href {http://arxiv.org/abs/gr-qc/0103083}
  {\path{arXiv:gr-qc/0103083}}.

\bibitem{Bilic_Tupper_Viollier_2002}
N.~{Bili{\'c}}, G.~B. {Tupper}, R.~D. {Viollier}, {Unification of dark matter
  and dark energy: the inhomogeneous Chaplygin gas}, Physics Letters B 535
  (2002) 17--21.
\newblock \href {http://arxiv.org/abs/astro-ph/0111325}
  {\path{arXiv:astro-ph/0111325}}.

\bibitem{Bento_Bertolami_Sen_2002}
M.~C. {Bento}, O.~{Bertolami}, A.~A. {Sen}, {Generalized Chaplygin gas,
  accelerated expansion, and dark-energy-matter unification}, \prd 66 (2002)
  043507.
\newblock \href {http://arxiv.org/abs/gr-qc/0202064}
  {\path{arXiv:gr-qc/0202064}}.

\bibitem{Bento_Bertolami_Sen_2003}
M.~C. {Bento}, O.~{Bertolami}, A.~A. {Sen}, {Generalized Chaplygin gas and
  cosmic microwave background radiation constraints}, \prd 67 (2003) 063003.
\newblock \href {http://arxiv.org/abs/astro-ph/0210468}
  {\path{arXiv:astro-ph/0210468}}.

\bibitem{Bean_Dore_2003}
R.~{Bean}, O.~{Dor{\'e}}, {Are Chaplygin gases serious contenders for the dark
  energy?}, \prd 68 (2003) 023515.
\newblock \href {http://arxiv.org/abs/astro-ph/0301308}
  {\path{arXiv:astro-ph/0301308}}.

\bibitem{Amendola_Finelli_Burigana_Carturan_2003}
L.~{Amendola}, F.~{Finelli}, C.~{Burigana}, D.~{Carturan}, {WMAP and the
  generalized Chaplygin gas}, \jcap 7 (2003) 005.
\newblock \href {http://arxiv.org/abs/astro-ph/0304325}
  {\path{arXiv:astro-ph/0304325}}.

\bibitem{Carturan_Finelli_2003}
D.~{Carturan}, F.~{Finelli}, {Cosmological effects of a class of fluid dark
  energy models}, \prd 68 (2003) 103501.
\newblock \href {http://arxiv.org/abs/astro-ph/0211626}
  {\path{arXiv:astro-ph/0211626}}.

\bibitem{Makler_deOliveira_Waga_2003}
M.~{Makler}, S.~Q. {de Oliveira}, I.~{Waga}, {Constraints on the generalized
  Chaplygin gas from supernovae observations}, Physics Letters B 555 (2003)
  1--6.
\newblock \href {http://arxiv.org/abs/astro-ph/0209486}
  {\path{arXiv:astro-ph/0209486}}.

\bibitem{Dev_Alcaniz_Jain_2003}
A.~{Dev}, J.~S. {Alcaniz}, D.~{Jain}, {Cosmological consequences of a Chaplygin
  gas dark energy}, \prd 67 (2003) 023515.
\newblock \href {http://arxiv.org/abs/astro-ph/0209379}
  {\path{arXiv:astro-ph/0209379}}.

\bibitem{Sandvik_Tegmark_Zaldarriaga_2004}
H.~B. {Sandvik}, M.~{Tegmark}, M.~{Zaldarriaga}, I.~{Waga}, {The end of unified
  dark matter?}, \prd 69 (2004) 123524.
\newblock \href {http://arxiv.org/abs/astro-ph/0212114}
  {\path{arXiv:astro-ph/0212114}}.

\bibitem{Hu_1998}
W.~{Hu}, Structure formation with generalized dark matter, \apj 506 (1998)
  485--494.
\newblock \href {http://arxiv.org/abs/astro-ph/9801234}
  {\path{arXiv:astro-ph/9801234}}.

\bibitem{Reis_Waga_Calvao_Joras_2003}
R.~R. {Reis}, I.~{Waga}, M.~O. {Calv{\~a}o}, S.~E. {Jor{\'a}s}, {Entropy
  perturbations in quartessence Chaplygin models}, \prd 68 (2003) 061302.
\newblock \href {http://arxiv.org/abs/astro-ph/0306004}
  {\path{arXiv:astro-ph/0306004}}.

\bibitem{Nozari_Azizi_Alipour_2011}
K.~{Nozari}, T.~{Azizi}, N.~{Alipour}, {Observational constraints on Chaplygin
  cosmology in a braneworld scenario with induced gravity and curvature
  effect}, \mnras 412 (2011) 2125--2136.
\newblock \href {http://arxiv.org/abs/1011.3395} {\path{arXiv:1011.3395}}.

\bibitem{Lamon_Woehr_2010}
R.~{Lamon}, A.~J. {W{\"o}hr}, {Quintessence and (anti-)Chaplygin gas in loop
  quantum cosmology}, \prd 81 (2010) 024026.
\newblock \href {http://arxiv.org/abs/0910.4891} {\path{arXiv:0910.4891}}.

\bibitem{Roy_Buchert_2010}
X.~{Roy}, T.~{Buchert}, {Chaplygin gas and effective description of
  inhomogeneous universe models in general relativity}, \cqg 27~(17) (2010)
  175013.
\newblock \href {http://arxiv.org/abs/0909.4155} {\path{arXiv:0909.4155}}.

\bibitem{Bento_Bertolami_Reboucas_Silva_2006}
M.~C. {Bento}, O.~{Bertolami}, M.~J. {Rebou{\c c}as}, P.~T. {Silva},
  {Generalized Chaplygin gas model, supernovae, and cosmic topology}, \prd 73
  (2006) 043504.
\newblock \href {http://arxiv.org/abs/gr-qc/0512158}
  {\path{arXiv:gr-qc/0512158}}.

\bibitem{Wu_Yu_2007}
P.~{Wu}, H.~{Yu}, {Constraints on the unified dark energy dark matter model
  from latest observational data}, \jcap 3 (2007) 015.
\newblock \href {http://arxiv.org/abs/astro-ph/0701446}
  {\path{arXiv:astro-ph/0701446}}.

\bibitem{Xu_Lu_2010}
L.~{Xu}, J.~{Lu}, {Cosmological constraints on generalized Chaplygin gas model:
  Markov Chain Monte Carlo approach}, \jcap 3 (2010) 025.
\newblock \href {http://arxiv.org/abs/1004.3344} {\path{arXiv:1004.3344}}.

\bibitem{Campos_Fabris_Perez_Piattella_2013}
J.~P. {Campos}, J.~C. {Fabris}, R.~{Perez}, O.~F. {Piattella}, H.~{Velten},
  {Does Chaplygin gas have salvation?}, European Physical Journal C 73 (2013)
  2357.
\newblock \href {http://arxiv.org/abs/1212.4136} {\path{arXiv:1212.4136}}.

\bibitem{Xu_Zhang_2016}
Y.-Y. {Xu}, X.~{Zhang}, {Comparison of dark energy models after Planck 2015},
  European Physical Journal C 76 (2016) 588.
\newblock \href {http://arxiv.org/abs/1607.06262} {\path{arXiv:1607.06262}}.

\bibitem{Freitas_Goncalves_Velten_2011}
R.~C. {Freitas}, S.~V.~B. {Gon{\c c}alves}, H.~E.~S. {Velten}, {Constraints on
  the generalized Chaplygin gas model from Gamma-ray bursts}, Physics Letters B
  703 (2011) 209--216.
\newblock \href {http://arxiv.org/abs/1004.5585} {\path{arXiv:1004.5585}}.

\bibitem{El_Zant_2015}
A.~A. {El-Zant}, {Unified dark matter: constraints from galaxies and clusters},
  \mnras 453 (2015) 2250--2258.
\newblock \href {http://arxiv.org/abs/1507.07369} {\path{arXiv:1507.07369}}.

\bibitem{Lu_Xu_Wu_Liu_2011}
J.~{Lu}, L.~{Xu}, Y.~{Wu}, M.~{Liu}, {Combined constraints on modified
  Chaplygin gas model from cosmological observed data: Markov Chain Monte Carlo
  approach}, General Relativity and Gravitation 43 (2011) 819--832.
\newblock \href {http://arxiv.org/abs/1105.1870} {\path{arXiv:1105.1870}}.

\bibitem{Paul_Thakur_Beesham_2014}
B.~C. {Paul}, P.~{Thakur}, A.~{Beesham}, {Observational constraints on Modified
  Chaplygin Gas from Large Scale Structure}, ArXiv e-prints\href
  {http://arxiv.org/abs/1410.6588} {\path{arXiv:1410.6588}}.

\bibitem{Zhang_Wu_Zhang_2006}
X.~{Zhang}, F.-Q. {Wu}, J.~{Zhang}, {New generalized Chaplygin gas as a scheme
  for unification of dark energy and dark matter}, \jcap 1 (2006) 003.
\newblock \href {http://arxiv.org/abs/astro-ph/0411221}
  {\path{arXiv:astro-ph/0411221}}.

\bibitem{Pourhassan_Kahya_2014}
B.~{Pourhassan}, E.~O. {Kahya}, {FRW cosmology with the extended Chaplygin
  gas}, Adv.~High Energy Phys. 2014 (2014) 231452.
\newblock \href {http://arxiv.org/abs/1405.0667} {\path{arXiv:1405.0667}}.

\bibitem{Carneiro_Pigozzo_2014}
S.~{Carneiro}, C.~{Pigozzo}, {Observational tests of non-adiabatic Chaplygin
  gas}, \jcap 10 (2014) 060.
\newblock \href {http://arxiv.org/abs/1407.7812} {\path{arXiv:1407.7812}}.

\bibitem{Marttens_etc_2017}
R.~F. {vom Marttens}, L.~{Casarini}, W.~{Zimdahl}, W.~S.
  {Hip{\'o}lito-Ricaldi}, D.~F. {Mota}, {Does a generalized Chaplygin gas
  correctly describe the cosmological dark sector?}, Physics of the Dark
  Universe 15 (2017) 114--124.
\newblock \href {http://arxiv.org/abs/1702.00651} {\path{arXiv:1702.00651}}.

\bibitem{Wen_Wang_2017}
S.~{Wen}, S.~{Wang}, {Comparing dark energy models with current observational
  data}, ArXiv e-prints\href {http://arxiv.org/abs/1708.03143}
  {\path{arXiv:1708.03143}}.

\bibitem{Bond_Efstathiou_Tegmark_1997}
J.~R. {Bond}, G.~{Efstathiou}, M.~{Tegmark}, {Forecasting cosmic parameter
  errors from microwave background anisotropy experiments}, \mnras 291 (1997)
  L33--L41.
\newblock \href {http://arxiv.org/abs/astro-ph/9702100}
  {\path{arXiv:astro-ph/9702100}}.

\bibitem{Efstathiou_Bond_1999}
G.~{Efstathiou}, J.~R. {Bond}, Cosmic confusion: degeneracies among
  cosmological parameters derived from measurements of microwave background
  anisotropies, \mnras 304 (1999) 75--97.
\newblock \href {http://arxiv.org/abs/astro-ph/9807103}
  {\path{arXiv:astro-ph/9807103}}.

\bibitem{Betoule_et_al_2014}
M.~{Betoule}, {et al.}, {Improved cosmological constraints from a joint
  analysis of the SDSS-II and SNLS supernova samples}, \aap 568 (2014) A22.
\newblock \href {http://arxiv.org/abs/1401.4064} {\path{arXiv:1401.4064}}.

\bibitem{Aurich_Lustig_2016}
R.~{Aurich}, S.~{Lustig}, {Early-matter-like dark energy and the cosmic
  microwave background}, \jcap 1 (2016) 021.
\newblock \href {http://arxiv.org/abs/1511.01691} {\path{arXiv:1511.01691}}.

\bibitem{Pettini_Cooke_2012}
M.~Pettini, R.~Cooke, A new, precise measurement of the primordial abundance of
  deuterium, \mnras 425 (2012) 2477.
\newblock \href {http://arxiv.org/abs/1205.3785} {\path{arXiv:1205.3785}}.

\bibitem{Dunkley_Bucher_Ferreira_Moodley_Skordis_2005}
J.~{Dunkley}, M.~{Bucher}, P.~G. {Ferreira}, K.~{Moodley}, C.~{Skordis}, {Fast
  and reliable Markov chain Monte Carlo technique for cosmological parameter
  estimation}, \mnras 356 (2005) 925--936.
\newblock \href {http://arxiv.org/abs/astro-ph/0405462}
  {\path{arXiv:astro-ph/0405462}}.

\bibitem{Suzuki_et_al_2012}
N.~{Suzuki}, {et. al}, {Supernova Cosmology Project}, {The Hubble Space
  Telescope Cluster Supernova Survey. V. Improving the Dark-energy Constraints
  above $z>1$ and Building an Early-type-hosted Supernova Sample}, \apj 746
  (2012) 85.
\newblock \href {http://arxiv.org/abs/1105.3470} {\path{arXiv:1105.3470}}.

\bibitem{Goliath_Amanullah_Astier_Goobar_Pain_2001}
M.~{Goliath}, R.~{Amanullah}, P.~{Astier}, A.~{Goobar}, R.~{Pain}, {Supernovae
  and the nature of the dark energy}, \aap 380 (2001) 6--18.
\newblock \href {http://arxiv.org/abs/astro-ph/0104009}
  {\path{arXiv:astro-ph/0104009}}.

\bibitem{Riess_et_al_2016}
A.~G. {Riess}, L.~M. {Macri}, S.~L. {Hoffmann}, D.~{Scolnic}, S.~{Casertano},
  A.~V. {Filippenko}, B.~E. {Tucker}, M.~J. {Reid}, D.~O. {Jones}, J.~M.
  {Silverman}, R.~{Chornock}, P.~{Challis}, W.~{Yuan}, P.~J. {Brown}, R.~J.
  {Foley}, {A 2.4\% Determination of the Local Value of the Hubble Constant},
  \apj 826 (2016) 56.
\newblock \href {http://arxiv.org/abs/1604.01424} {\path{arXiv:1604.01424}}.

\bibitem{Planck_2015_I}
{Planck Collaboration}, R.~{Adam}, P.~A.~R. {Ade}, N.~{Aghanim}, Y.~{Akrami},
  M.~I.~R. {Alves}, M.~{Arnaud}, F.~{Arroja}, J.~{Aumont}, C.~{Baccigalupi},
  et~al., {Planck 2015 results. I. Overview of products and scientific
  results}, \aap 594 (2016) A1.
\newblock \href {http://arxiv.org/abs/1502.01582} {\path{arXiv:1502.01582}}.

\bibitem{Eisenstein_Hu_1998}
D.~J. {Eisenstein}, W.~{Hu}, {Baryonic Features in the Matter Transfer
  Function}, \apj 496 (1998) 605--614.
\newblock \href {http://arxiv.org/abs/astro-ph/9709112}
  {\path{arXiv:astro-ph/9709112}}.

\bibitem{Aubourg_etc_2015}
{\'E}.~{Aubourg}, {et al.}, {BOSS Collaboration}, {Cosmological implications of
  baryon acoustic oscillation measurements}, \prd 92 (2015) 123516.
\newblock \href {http://arxiv.org/abs/1411.1074} {\path{arXiv:1411.1074}}.

\bibitem{Hu_Sugiyama_1996}
W.~{Hu}, N.~{Sugiyama}, {Small-Scale Cosmological Perturbations: an Analytic
  Approach}, \apj 471 (1996) 542--570.
\newblock \href {http://arxiv.org/abs/astro-ph/9510117}
  {\path{arXiv:astro-ph/9510117}}.

\bibitem{Beutler_et_al_2011}
F.~{Beutler}, C.~{Blake}, M.~{Colless}, D.~H. {Jones}, L.~{Staveley-Smith},
  L.~{Campbell}, Q.~{Parker}, W.~{Saunders}, F.~{Watson}, {The 6dF Galaxy
  Survey: baryon acoustic oscillations and the local Hubble constant}, \mnras
  416 (2011) 3017--3032.
\newblock \href {http://arxiv.org/abs/1106.3366} {\path{arXiv:1106.3366}}.

\bibitem{Ross_et_al_2015}
A.~J. {Ross}, L.~{Samushia}, C.~{Howlett}, W.~J. {Percival}, A.~{Burden},
  M.~{Manera}, {The clustering of the SDSS DR7 main Galaxy sample - I. A 4 per
  cent distance measure at z = 0.15}, \mnras 449 (2015) 835--847.
\newblock \href {http://arxiv.org/abs/1409.3242} {\path{arXiv:1409.3242}}.

\bibitem{Chuang_et_al_2016}
C.-H. {Chuang}, {et. al}, {The Clustering of Galaxies in the Completed SDSS-III
  Baryon Oscillation Spectroscopic Survey: single-probe measurements from DR12
  galaxy clustering -- towards an accurate model}, \mnras 471 (2017)
  2370--2390.
\newblock \href {http://arxiv.org/abs/1607.03151} {\path{arXiv:1607.03151}}.

\bibitem{Delubac_et_al_2015}
T.~{Delubac}, {et al.}, {Baryon acoustic oscillations in the Ly{$\alpha$}
  forest of BOSS DR11 quasars}, \aap 574 (2015) A59.
\newblock \href {http://arxiv.org/abs/1404.1801} {\path{arXiv:1404.1801}}.

\bibitem{Font_Ribera_et_al_2014}
A.~{Font-Ribera}, {et al.}, {Quasar-Lyman {$\alpha$} forest cross-correlation
  from BOSS DR11: Baryon Acoustic Oscillations}, \jcap 5 (2014) 027.
\newblock \href {http://arxiv.org/abs/1311.1767} {\path{arXiv:1311.1767}}.

\bibitem{Bautista_et_al_2017}
J.~E. {Bautista}, {et al.}, {Measurement of BAO correlations at $z=2.3$ with
  SDSS DR12 Lya-Forests}, \aap 603 (2017) A12.
\newblock \href {http://arxiv.org/abs/1702.00176} {\path{arXiv:1702.00176}}.

\bibitem{Moresco_et_al_2016}
M.~{Moresco}, L.~{Pozzetti}, A.~{Cimatti}, R.~{Jimenez}, C.~{Maraston},
  L.~{Verde}, D.~{Thomas}, A.~{Citro}, R.~{Tojeiro}, D.~{Wilkinson}, {A 6\%
  measurement of the Hubble parameter at z\~{}0.45: direct evidence of the
  epoch of cosmic re-acceleration}, \jcap 5 (2016) 014.
\newblock \href {http://arxiv.org/abs/1601.01701} {\path{arXiv:1601.01701}}.

\end{thebibliography}

\bibliographystyle{elsarticle-num}

\end{document}